\documentclass[aps,prd,twocolumn,showpacs,superscriptaddress,groupedaddress]{revtex4}
\usepackage{ads}
\usepackage{jjzhang}
\usepackage{xcolor}
\usepackage{graphicx,color,bm}

\usepackage{amsmath,amssymb}
\usepackage[colorlinks,linkcolor=blue,citecolor=blue]{hyperref}
\usepackage{epstopdf}
\usepackage{ulem}
\usepackage{array}
\usepackage{verbatim}
\usepackage[utf8]{inputenc}
\usepackage{rotating}
\newcolumntype{K}[1]{>{\centering\arraybackslash}m{#1}}

\begin{document}

\title{A Fully Self-Consistent Cosmological Simulation Pipeline for Interacting Dark Energy Models}

\author{Jiajun Zhang\footnote{liamzhang@sjtu.edu.cn}}
\affiliation{Department of Astronomy, School of Physics and Astronomy, Shanghai Jiao Tong University, Shanghai, 200240, China}
\author{Rui An\footnote{{an\_rui@sjtu.edu.cn}}}
\affiliation{Department of Astronomy, School of Physics and Astronomy, Shanghai Jiao Tong University, Shanghai, 200240, China}
\affiliation{IFSA Collaborative Innovation Center, Shanghai Jiao Tong University, Shanghai, China}
\author{Shihong Liao}
\affiliation{Key Laboratory for Computational Astrophysics, National Astronomical Observatories, Chinese Academy of Sciences, Beijing, 100012, China}
\author{Wentao Luo}
\affiliation{Department of Astronomy, School of Physics and Astronomy, Shanghai Jiao Tong University, Shanghai, 200240, China}
\affiliation{Kavli Institute for the Physics and Mathematics of the Universe (Kavli IPMU, WPI),
Tokyo Institutes for Advanced Study, The University of Tokyo, Chiba, 277-8582, Japan}
\author{Zhaozhou Li}
\affiliation{Department of Astronomy, School of Physics and Astronomy, Shanghai Jiao Tong University, Shanghai, 200240, China}
\author{Bin Wang \footnote{{wangb@yzu.edu.cn}}}
\affiliation{Center for Gravitation and Cosmology, College of Physical Science and Technology, Yangzhou University, Yangzhou, 225009, China}
\affiliation{Department of Astronomy, School of Physics and Astronomy, Shanghai Jiao Tong University, Shanghai, 200240, China}

\begin{abstract}
We devise a fully self-consistent simulation pipeline for the first time to study the interaction between dark matter and dark energy.  We perform convergence tests and show that our code is accurate on different scales. Using the parameters constrained by Planck, Type Ia Supernovae, Baryon Acoustic Oscillations (BAO) and Hubble constant observations, we  perform cosmological N-body simulations. We calculate the resulting matter power spectra and halo mass functions for four different interacting dark energy models. In addition to the dark matter density distribution, we also show the inhomogeneous density distribution of dark energy. With this new simulation pipeline, we can further refine and constrain interacting dark energy models.
\end{abstract}
\date{\today}

\maketitle
\section{Introduction}

It is widely believed that our Universe is undergoing an accelerated expansion. Within the framework of Einstein gravity, this acceleration can be driven by a new energy component with negative pressure, called dark energy. In the standard $\Lambda$-cold dark matter ($\Lambda$CDM) model, this mysterious energy is explained as the cosmological constant, $\Lambda$. The standard model is commonly used to describe the evolution of the Universe and it is consistent with a number of observations. However, from the theoretical point of view, the $\Lambda$CDM model faces significant challenges such as the cosmological constant problem \cite{Weinberg1989} and the coincidence problem \cite{Zlatev1999}. Recently, people have found inconsistencies when comparing different observations assuming $\Lambda$CDM model. These include, i) a $\sim3\sigma$ mismatch between the Hubble constant inferred from the cosmic microwave background (CMB) measurements and that from the direct local observations \cite{Riess2011,Riess2016}, ii) a $\sim2.5\sigma$ discrepancy between the Hubble parameter and angular distance at $z=2.34$ measured from the Baryon Oscillation Spectroscopic Survey (BOSS) experiment and that inferred from the CMB measurements \cite{Delubac2015}, iii) a $\sim2.3\sigma$ tension between the weak lensing data taken from a $450-\text{deg}^2$ observing field of the Kilo Degree Survey (KiDS) and the Planck 2015 CMB data \cite{Hildebrandt2017}. All of these theoretical and observational challenges clearly indicate the need to investigate alternative cosmological models. 

Given the fact that the Universe is composed of nearly $25\%$ dark matter (DM) and $70\%$ dark energy (DE) today, it is natural to ask whether these two most abundant components of the Universe can interacte with each other instead of evolving separately. It was reported that appropriate interactions between DM and DE can provide a mechanism to alleviate the coincidence problem  \cite{Amendola2000,Amendola2003,Amendola2006,Pavon2005,Boeher2008,Olivares2006,Chen2008}. It was shown that the interacting DM and DE (IDE here after in short) models are consistent with CMB observations, and they are able to relieve the discordance between BOSS and CMB measurements mentioned before \cite{Ferreira2014}. Moreover, it was shown that the IDE models can alleviate the tension between weak lensing and CMB measurements \cite{An2018}. Since the nature of neither DM nor DE is known, mostly phenomenological model for interactions between them have been studied (see \cite{Wang2016} for a recent review and references therein). Quantum field theory of dark energy interacting with dark matter was recently discussed in \cite{amico2016prd,marsh2017prl}. 

N-body simulations have been widely adopted to study the non-linear evolution of the large scale structure of the Universe. Because the IDE model is different from the $\Lambda$CDM model in  many aspects, it is important to build up a fully self-consistent simulation pipeline to study the non-linear structure formation in IDE models. Some attempts were made to build simulation codes for IDE models\citep{baldi2010mnras,baldi2011mnras}. However, the inputs used in such codes were not self-consistent. For example the initial power spectrum was generated assuming $\Lambda$CDM model. Moreover, a simplified DE distribution that was constant in different scales and redshifts was used. 

In this paper, we propose a fully self-consistent simulation pipeline for general phenomenological IDE models. We do not limit the DE to be in the quintessence region $-1<w_d<-1/3$, but allow its equation of state to be either bigger or smaller than $-1$. We include the DE perturbation by self-consistently solving its linear level perturbation equations. All initial conditions we put in the simulation use the parameters constrained by observations for IDE models. We find that the non-linear structure formation at low redshift can put further constraints on IDE models. 

The organization of the paper is as following. We first introduce our phenomenological IDE models and the simulation pipeline in Sec.~\ref{sec:method}. The details about the design of the simulation, the comparison with previous works and the code convergence tests are given in Sec.~\ref{sec:sim}. Then we show the main results including halo mass functions and non-linear matter power spectra of the models in Sec.~\ref{sec:result}. Finally, we summarize and discuss our results in Sec.~\ref{sec:conclusion}. 

\section{Methodology}\label{sec:method}
\subsection{Phenomenological Model}

We consider a phenomenological IDE model which has been widely discussed \cite{Wang2016}. In this model, the covariant description of the energy-momentum transfer between DE and DM is given by
\begin{equation}
\label{eq.Tdmde}
\bigtriangledown_{\mu}T^{\mu\nu}_{(\lambda)}=Q^{\nu}_{(\lambda)},
\end{equation}
where $Q^{\nu}$ denotes the interaction between two dark components and $\lambda$ denotes either the DM or the DE sector. For the whole system, the energy momentum conservation still holds, satisfying
\begin{equation}
\label{eq.Ttotal}
\sum_{\lambda}\bigtriangledown_{\mu}T^{\mu\nu}_{(\lambda)}=0.
\end{equation}
Here we work with the general stress-energy tensor 
\begin{equation}
\label{eq.T}
T^{\mu\nu}=\rho U^{\mu}U^{\nu}+p(g^{\mu\nu}+U^{\mu}U^{\nu}).
\end{equation}
The zero-component of Eq. (\ref{eq.Tdmde}) provides the background conservation equations for the energy densities of the dark sectors,
\begin{gather}
\label{eq.rhodm}
\rho'_{c}+3\mathcal{H}\rho_{c}=a^2Q_{(c)}^0=Q,\\
\label{eq.rhode}
\rho'_{d}+3\mathcal{H}(1+w_d)\rho_{d}=-a^2Q_{(d)}^0=-Q,
\end{gather}
where the subscript ``c" denotes DM and ``d" denotes DE. $\mathcal{H}$ is the Hubble function defined as $\mathcal{H}=a'/a$, $a$ is the cosmic scale factor and the prime is the derivative with respect to the conformal time, and $w_d=p_d/\rho_d$ is the constant equation of state for DE. $Q$ represents the interaction kernel, which is written as a  linear combination of the energy densities of dark sectors in the form of $Q=3\xi_{1}\mathcal{H}\rho_{c}+3\xi_{2}\mathcal{H}\rho_{d}$, where $\xi_1$ and $\xi_2$ are free parameters to be determined from observations. $Q>0$ indicates the energy flows from DE to DM while $Q<0$ signals the opposite. In Table \ref{tab.model} we list four phenomenological interacting models explored in this work. We study the constant equation of state of DE in the phantom and quintessence regions, respectively, to ensure stable density perturbations \cite{He2008}.

\begin{table}[ht]
\caption{Phenomenological interacting models \label{tab.model}}
\begin{tabular}{ccc}
\hline
Model & $Q$ & $w_d$\\
\hline
I & $3\xi_{2}\mathcal{H}\rho_{d}$ & $-1<w_d<-1/3$ \\
II & $3\xi_{2}\mathcal{H}\rho_{d}$ & $w_d<-1$ \\
III & $3\xi_{1}\mathcal{H}\rho_{c}$ & $w_d<-1$ \\
IV & $3\xi\mathcal{H}(\rho_{c}+\rho_{d})$ & $w_d<-1$ \\
\hline
\end{tabular}
\end{table}

The perturbed space-time is given by
\begin{align}
\label{eq.ds}
ds^2=&a^2(\tau)[-(1+2\psi)d\tau^2+2\partial_iBd\tau dx^i \notag \\
&+(1+2\phi)\delta_{ij}dx^idx^j+D_{ij}Edx^idx^j],
\end{align}
where $\psi$, $B$, $\phi$, and $E$ represent the scalar metric perturbations, and $D_{ij}=(\partial_i\partial_j-\frac{1}{3}\delta_{ij})\bigtriangledown^2$. 

The linear perturbation equations of IDE models were derived in \cite{He2008,He2009}. 
The gauge invariant gravitational potentials, density contrast, and peculiar velocity are described as follows,
\begin{align}
&\Psi=\psi-\frac{1}{k}\mathcal{H}(B+\frac{E'}{2k})-\frac{1}{k}(B'+\frac{E''}{2k}), \\
&\Phi=\phi+\frac{1}{6}E-\frac{1}{k}\mathcal{H}(B+\frac{E'}{2k}), \\
&D_{\lambda}=\delta_{\lambda}-\frac{\rho'_{\lambda}}{\rho_{\lambda\mathcal{H}}}(\phi+\frac{E}{6}),\\
&V_{\lambda}=v_{\lambda}-\frac{E'}{2k}.
\end{align}
Choosing the Longitudinal gauge by defining $E=0$, $B=0$, we have
\begin{align}
&\Psi=\psi, \\
&\Phi=\phi, \\
&D_{\lambda}=\delta_{\lambda}-\frac{\rho'_{\lambda}}{\rho_{\lambda}\mathcal{H}}\Phi,\\
&V_{\lambda}=v_{\lambda}.
\end{align}
Considering the phenomenological form of the energy transfer between dark sectors defined above, we obtain the general gauge invariant perturbation equations for DM and DE respectively,
\begin{subequations}
\begin{align}
\label{eq.DUdm}
D'_c=&-kU_c+6\mathcal{H}\Psi(\xi_1+\xi_2/r)-3(\xi_1+\xi_2/r)\Phi' \notag \\
&+3\mathcal{H}\xi_2(D_d-D_c)/r,  \\
U'_c=&-\mathcal{H}U_c+k\Psi-3\mathcal{H}(\xi_1+\xi_2/r)U_c, \notag
\end{align}
\begin{align}
\label{eq.DUde}
D'_d=&-3\mathcal(C_e^2-w_d)D_d-9\mathcal{H}^2(C_e^2-C_a^2)\frac{U_d}{k} \notag \\
&+[3w'_d-9\mathcal{H}(w_d-C_e^2)(\xi_1r+\xi_2+1+w_d)]\Phi \notag \\
&+3(\xi_1r+\xi_2)\Phi'-3\Psi\mathcal{H}(\xi_1r+\xi_2) \notag \\
&-9\mathcal{H}^2(C_e^2-C_a^2)(\xi_ir+\xi_2)\frac{U_d}{(1+w_d)k} \notag \\
&-kU_d+3\mathcal{H}\xi_1r(D_d-D_c), \\
U'_d=&-\mathcal{H}(1-3w_d)U_d+3(C_e^2-C_a^2)\mathcal{H}U_d \notag \\
&-3kC_e^2(\xi_1r+\xi_2+1+w_d)\Phi+kC_e^2D_d \notag \\
&+3\mathcal{H}(C_e^2-C_a^2)(\xi_1r+\xi_2)\frac{U_d}{1+w_d} \notag \\
&+(1+w_d)k\Psi+3\mathcal{H}(\xi_1r+\xi_2)U_d, \notag
\end{align}
\end{subequations}
where $U_{\lambda}=(1+w_{\lambda})V_{\lambda}$, $C_e^2$ is the effective sound speed of DE, $C_a^2$ is the adiabatic sound speed, and $r=\rho_c/\rho_d$ is the energy density ratio of DM and DE.

From the perturbed Einstein equations, we can get the Poisson equation in the subhorizon approximation \cite{He2009}
\begin{equation}
\label{eq.Poisson}
-k^2\Psi=\frac{3}{2}\mathcal{H}^2[\Omega_c\bigtriangleup_c+(1-\Omega_c)\bigtriangleup_d],
\end{equation}
where $\bigtriangleup_{\lambda}=\delta_{\lambda}-\frac{\rho'_{\lambda}}{\rho_{\lambda}}\frac{V_{\lambda}}{k}$, $\Omega_{\lambda}=\frac{\rho_{\lambda}}{\rho_{crit}}$, and $\rho_{crit}$ is the critical density. This equation can be used to build the bridge between the matter perturbations and the metric perturbations. We can rewrite the Poisson equation in real space as
\begin{equation}
\label{eq.RPossion}
\bigtriangledown^2\Psi=-\frac{3}{2}\mathcal{H}^2[\Omega_c\bigtriangleup_c+(1-\Omega_c)\bigtriangleup_d].
\end{equation}
The second equation in (\ref{eq.DUdm}) can give the velocity perturbation for DM of the form
\begin{equation}
\label{eq.vdm}
V'_c+[\mathcal{H}+3\mathcal{H}(\xi_1+\xi_2/r)]V_c-k\Psi=0.
\end{equation}
Combining the Poisson equation (\ref{eq.RPossion}), this equation can be written in real space and in terms of the effective gravitational potential to give a modified Euler equation
\begin{align}
\label{eq.Euler}
&\bigtriangledown V'_c+[\mathcal{H}+3\mathcal{H}(\xi_1+\xi_2/r)]\bigtriangledown V_c \notag \\
&+\frac{3}{2}\mathcal{H}^2[\Omega_c\bigtriangleup_c+(1-\Omega_c)\bigtriangleup_d]=0.
\end{align}
It is clear from the above equation that due to the coupling between dark sectors, the gravitational potential is modified and there is an additional acceleration for DM particles. 

 The four phenomenological interacting models listed in Table \ref{tab.model} were recently investigated by \cite{costa2017jcap} to constrain them by employing recent observational data sets including CMB data from Planck 2015, Type Ia supernovae (SNIa), baryon acoustic oscillations (BAO), Hubble constant (H0). We use their numerical fitting results as the input parameters for our simulations and investigate the effects of the interaction between dark sectors on the structure formation by performing N-body simulations. We use the Planck 2015 parameters \cite{Ade2016} for the fiducial $\Lambda$CDM model to compare our results from the IDE models. The cosmological parameters we use in our computations are listed in Table \ref{tab.parameters}.

\begin{table*}
\caption{Cosmological parameters (PBSH=Planck+BAO+SNIa+H0) \label{tab.parameters}}
\begin{tabular}{cccccccccc}
\hline
    & \multicolumn{2}{c}{IDE\_I} & \multicolumn{2}{c}{IDE\_II} & \multicolumn{2}{c}{IDE\_III} & \multicolumn{2}{c}{IDE\_IV} & \multicolumn{1}{c}{$\Lambda$CDM} \\
    \cline{2-10}
    \ \ Parameter\ \  &\  \ Planck\ \ &\ \  PBSH \ \ &Planck\ \  &\ \  PBSH\ \  &\ \  Planck \ \ &\ \  PBSH\ \  & \ \ Planck \ \  &\ \ PBSH\ \  & \ \ Planck\ \  \\
    \hline
$\Omega_b h^2$ & 0.0222 & 0.02223 & 0.02225 & 0.02224 & 0.02235 & 0.02228 & 0.02235 &  0.02228 & 0.02225\\
$\Omega_c h^2$ & 0.07131 & 0.0792 & 0.1334 & 0.1351 & 0.1236 & 0.1216 & 0.124 & 0.1218 & 0.1198\\
$100\theta_{MC}$ &1.044 & 1.043 & 1.04 & 1.04 &1.041 & 1.041 & 1.041 & 1.041 & 1.04077\\
$\tau$ & 0.08063 & 0.08204 & 0.07653 & 0.081 & 0.07051 & 0.07728 & 0.07043 & 0.07709 & 0.079\\
${\rm{ln}}(10^{10} A_s)$ & 3.097 & 3.099 & 3.088 & 3.097 & 3.074 & 3.088 & 3.073 & 3.087 & 3.094\\
$n_s$ & 0.9633 & 0.9645 & 0.9638 & 0.9643 & 0.9608 & 0.9624 & 0.9609 & 0.9624 & 0.9645\\
$w_d$ & -0.9031& -0.9191 & -1.55 & -1.088 & -1.702 & -1.104 & -1.691 & -1.105 & -1 \\
$\xi_1$ & -- & -- & -- & -- & 0.001458 & 0.0007127 & 0.001416 & 0.000735 & -- \\
$\xi_2$ &-0.1297 & -0.1107 & 0.03884 & 0.05219 & -- & -- & 0.001416 & 0.000735 & --\\
\hline
$H_0$ & 68.1 & 68.18 & 83.88 & 68.35 & 84.91 & 68.91 & 84.63 & 68.88 &67.27 \\
$\Omega_m$ & 0.2101 & 0.2204 & 0.2312 & 0.3384 & 0.212 & 0.3045 & 0.2141 & 0.3053 & 0.3156\\
\hline
\end{tabular}
\end{table*}

\subsection{Initial Condition}
We use the capacity constrained Voronoi tessellation (CCVT) method, which is an alternative method to produce a uniform and isotropic particle distribution, to generate pre-initial conditions \citep{liao2018ccvt}. In comparison to the gravitational equilibrium state (glass \citep{white1996glass}), the CCVT configuration is a geometrical equilibrium state, and is a more natural choice for models that include forces other than pure gravity. We use the open source code 2LPTic \citep{2lptic} to generate the initial condition for all our simulations. We have modified 2LPTic such that it can read matter power spectra generated by CAMB\citep{camb} at arbitrary redshifts. Additional modifications were done such that it can read the $H(a)$ table and $m(a)$ table shown in Fig.~\ref{fig:interpolation}. As IDE models modify the Hubble diagram and matter density, this modification allows the code to use a consistent second-order Lagrangian Perturbation Theory.

In 2LPTic \citep{2lptic}, the second-order Lagrangian Perturbation Theory is applied following the equation describing the position displacement,
\begin{equation}
\mathbf{x(q)}=\mathbf{q}+\nabla_q\Psi^{(1)}+\nabla_q\Psi^{(2)},
\end{equation}
where $\Psi^{(1)}$ and $\Psi^{(2)}$ are first and second order displacement field, respectively.
The velocity displacement is given by
\begin{equation}
\mathbf{v(q)}=f_1H\nabla_q\Psi^{(1)}+f_2H\nabla_q\Psi^{(2)},
\end{equation}
where 
\begin{equation}\label{f1f2}
f_1=\dfrac{dln(D_1)}{dlna},\\
f_2=\dfrac{dln(D_2)}{dlna}.
\end{equation}
We know $\Omega_m$ and $H$ in IDE models are different from those in the $\Lambda$CDM model. In the original 2LPTic code, $f_1\approx\Omega_m^{3/5}$ and $f_2\approx 2\Omega_m^{4/7}$ are calculated from the $\Lambda$CDM model \citep{scoccimarro1998mnras,crocce2006mnras}. We modified 2LPTic code such that it can read in the values of $f_1$, $f_2$, $\Omega_m$, and $H$ at arbitrary redshift calculated by the modified CAMB for our IDE models \citep{costa2017jcap}. However, at high redshift such as $z=49$ as used in our simulations, we find that $f_1\approx\Omega_m^{3/5}$ and $f_2\approx 2\Omega_m^{4/7}$ are very good approximations even for our IDE models. Thus for simplicity, we use this approximation to calculate $f_1$ and $f_2$ instead of using the results from the modified CAMB. We note however, that values of $f_1$ and $f_2$ calculated from the modified CAMB can be easily used in our simulations.

\subsection{ME-Gadget Code}
In order to study the IDE models, there are four modifications required in the cosmological simulations compared to the $\Lambda$CDM simulations \citep{baldi2010mnras,baldi2011mnras}. First of all, since the expansion of the universe is different in IDE models, the Hubble diagram $H(a)$ should be explicitly given. Secondly, because of the energy flow between DM and DE, the mass of the simulation particles $m(a)$, which represents the DM energy density, should be changed as a function of scale factor. Thirdly, the DM particles in the simulation will receive an additional acceleration proportional to its velocity $\mathbf{a_v}=\alpha (a)\mathbf{v}$, where 
\begin{equation}\label{eq:alphaa}
\alpha (a)=-3\mathcal{H}(\xi_1+\xi_2/r)a.    
\end{equation} Compared to Eq.(\ref{eq.Euler}), there is an additional minus sign and scale factor $a$, which come from the coordinate transformation \cite{baldi2010mnras}. $\mathbf{a_v}$ is referred to as the dragging force or friction term, although it is not necessarily slowing down the particles.

Finally, the gravitational constant $G$ is different from $\Lambda$CDM model. As a result, the DM particles in the simulations will experience an additional force, which is also called the fifth force. In fact, from Eq. (\ref{eq.Poisson}), we can see that the fifth force is caused by the perturbation of DE. Therefore, the fifth force is a modification to the Poisson equation in harmonic space $-k^2\Psi=\frac{3}{2}\mathcal{H}^2 \Omega_c\bigtriangleup_c(1+\beta(a,k))$, where $\beta(a,k)=(1-\Omega_c)\bigtriangleup_d /\Omega_c\bigtriangleup_c$. In \citep{baldi2010mnras,baldi2011mnras}, $\beta(a,k)$ was simplified to be a constant. This however, is not accurate enough for capturing the distributions of DE and DM. In contrast, we use $\beta(a,k)$ as a two-dimensional function, which is calculated by the modified CAMB. We applied the above four modifications in the original N-body simulation code Gadget2 \citep{gadget2}, and named it ME-Gadget.

In order to implement the four modifications in $H(a),m(a),\alpha(a),$ and $\beta(a,k)$, we first make tables at discrete values of $a$ and $k$. We use one dimensional cubic interpolation for $H(a),m(a)$ and $\alpha(a)$, two dimensional bilinear interpolation for $\beta(a,k)$. At every time step, $H(a)$ is used to calculate the length of the time step, $m(a)$ is used to update the mass of the simulation particle, and $\alpha(a)$ is used to update the velocity of the simulation particle together with the acceleration from gravity. 

The use of $\beta(a,k)$ in our code is explained below in detail. In every time step, when the code calculates the particle-mesh gravity force, it will perform Fourier transform and solve the Poisson equation in harmonic space. At this time, the gravitational potential field in harmonic space is calculated. $\beta(a,k)$ is used to modify this gravitational potential field according to $a$ and $k$. We assume that the DE perturbation is only effective at large scales, and thus at small scales, gravity follows the normal Poisson equation. Therefore, only modifying the particle-mesh gravity solver, which solves the gravity in long range part, is accurate enough. By the same argument, the $\beta(a,k)$ we adopted in the simulation is calculated by our modified CAMB. $\beta(a,k)$ is the ratio between the linear perturbation of DE and the linear perturbation of DM. This treatment is different from \citep{baldi2010mnras,baldi2011mnras}. Although, the linear calculated DE perturbation has limitations, we use it here for lack of better choice. Besides, the effect of $\beta(a,k)$ is minor compared to $m(a)$ and $\alpha(a)$. The lowest panels of Fig.~\ref{fig:interpolation} show the values of $\beta(a,k)$ at $a=1$ as a function of $k$ for linear perturbations. We can see clearly that the argument that DE perturbation is only effective at large scales is correct, and it is orders of magnitude smaller than DM perturbation. Gadget2 code calculates the gravitational force using the tree algorithm at short range. The tree force part however, is not modified since it solves the short range force in our code. We have tested that loading the data from the tables and performing interpolations do not affect the code efficiency significantly. For the same resolution and box size, a $\Lambda$CDM simulation and an IDE simulation cost almost the same time with the same number of CPU cores. We note that our code can also easily handle other non-standard cosmological models by simply modifying the input tables.  

\section{Simulation}\label{sec:sim}

\subsection{Simulation Sets}
We have run three different series of simulations that include comparison test runs, convergence test runs, and scientific runs. 

For the comparison test runs, we simulate models studied by Ref.~\citep{baldi2011mnras} using our code and compare the results. Comparing the non-linear power spectrum at $z=0$, we find that our results are consistent. These runs are named as $S0,\ S1,\ S2,\ S3,\ S4,$ and $S6$ following the naming convention in Ref.\citep{baldi2011mnras}.

For the convergence test runs, we study the effects of the number of grids, the box size, and the resolution. The effect of number of grids on the final results needs to be tested for three reasons; a) we need to make sure whether neglecting the modification of short-range force is a valid choice, b) whether assuming DE perturbation is effective only at long range, is reasonable, c) the number of grids is sufficient to capture the large scale structure we would like to examine. Along with the number of grids, we checked the usual effects of changing box size and resolution. This is done to check the systematic uncertainty of the simulations. In addition, it will be useful for future studies with different box size and resolution. Thus, we perform convergence test runs and compare the non-linear power spectrum and halo mass function measured from simulations with different number of mesh grids, box sizes and resolution. The names and parameters of the simulations for convergence tests are summarized in Table~\ref{tab:convergence}.
\begin{table}
\caption{Simulations for convergence tests}\label{tab:convergence}
\begin{center}
\begin{tabular}{cccc}
\hline
 Name&Box size/$h^{-1}\mathrm{Mpc}$&$N_\text{particles}$&PMGRID\\
\hline
 PM128&400&$256^3$&128\\
 PM256&400&$256^3$&256\\
 PM512&400&$256^3$&512\\
 BOX&800&$512^3$&512\\
 RES&400&$512^3$&256\\
\hline
\end{tabular}
\end{center}
\end{table}

We perform the scientific runs using the parameters constrained by Planck CMB observation only and the combined Planck+BAO+SNIa+H0 constraints obtained in \citep{costa2017jcap}. In all these runs, we use a box size of 400 $h^{-1}$Mpc, $256^3$ particles, and $256$ mesh grids per dimension. The parameters we used are summarized in Table~\ref{tab.parameters}.

\subsection{Comparison Test}
\begin{figure}

\includegraphics[width=0.5\textwidth]{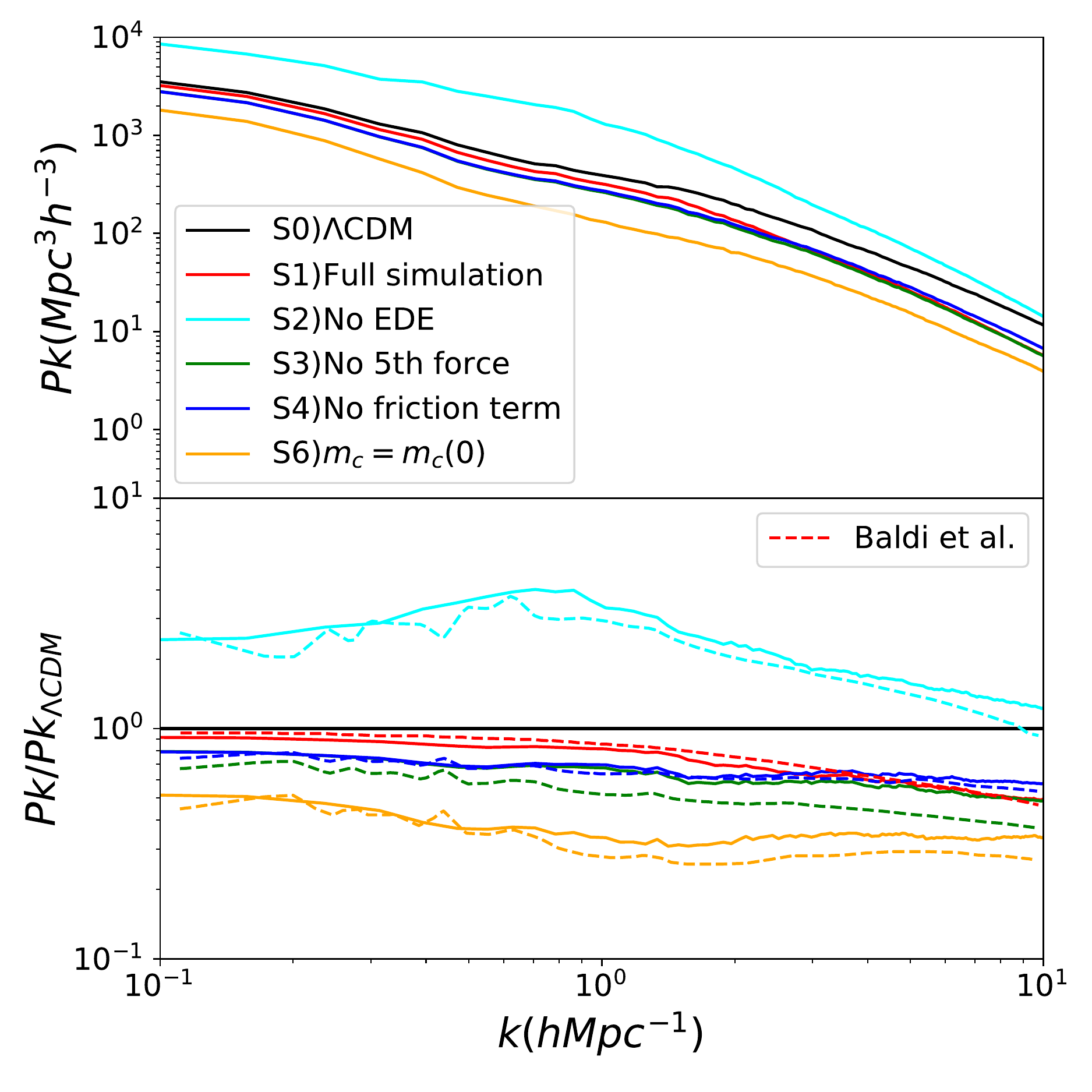}
\caption{The matter power spectra at $z=0$ measured from our simulations. S0, S1, S2, S3, S4 and S6 share the same convention in \cite{baldi2011mnras}. The upper panel plots the power spectra from different simulations, and the lower panel gives their ratios with respect to the $\Lambda$CDM one. The solid lines present our simulation results, while the dashed lines show those from \cite{baldi2011mnras}.
Qualitatively, our results are consistent with \cite{baldi2011mnras}. Notice that the red lines represent the final comparison of our simulation with \citep{baldi2011mnras}, which are quite similar.}\label{fig:baldi_compare}
\end{figure}

We have used the same model and the same parameters as \cite{baldi2011mnras} to test the performance of the ME-Gadget code. We have performed S0, S1, S2, S3, S4 and S6 simulations described in \cite{baldi2011mnras}. As shown in Fig.~\ref{fig:baldi_compare}, we find that our simulation results are consistent with \cite{baldi2011mnras} qualitatively for all the simulations including S1, which incorporates all the modifications. Comparing S1 and S2, we can see that modified Hubble diagram suppresses the matter power spectrum at all scales. Comparing S1 and S3, it is clear that modified fifth force, or the DE perturbation enhances the matter power spectrum at all scales. Comparing S1 and S4, we can see that modified velocity dependent acceleration (friction term or dragging force) suppresses the matter power spectrum mainly at small scales. Finally, comparing S1 and S6, it is clear that modified simulation particle mass enhances the matter power spectrum at all scales.

We have also tested that with further modifications of the tree force as \cite{baldi2011mnras} i.e., change the gravitational constant when computing the gravitational force in the tree algorithm. The difference is negligible at large scales ($<1\%$) and is also not very significant at small scales ($\sim5\%$). However, we do not plot the results of this tree force modifying test in Fig. 1 for a better illustration. 

Even though our results qualitatively agree with \cite{baldi2011mnras}, quantitatively, they are not exactly the same. The main reason is that our simulations are N-body simulations without gas, while \cite{baldi2011mnras} also included gas hydrodynamics. However the qualitatively agreement is sufficient to confirm that our implementation of the algorithm proposed in \cite{baldi2010mnras} is correct. We stress that comparing the quantitative results is not the main purpose of this work, and studying the effect of gas is beyond the scope of this paper.

\subsection{Convergence Test}
\begin{figure}
\includegraphics[width=0.5\textwidth]{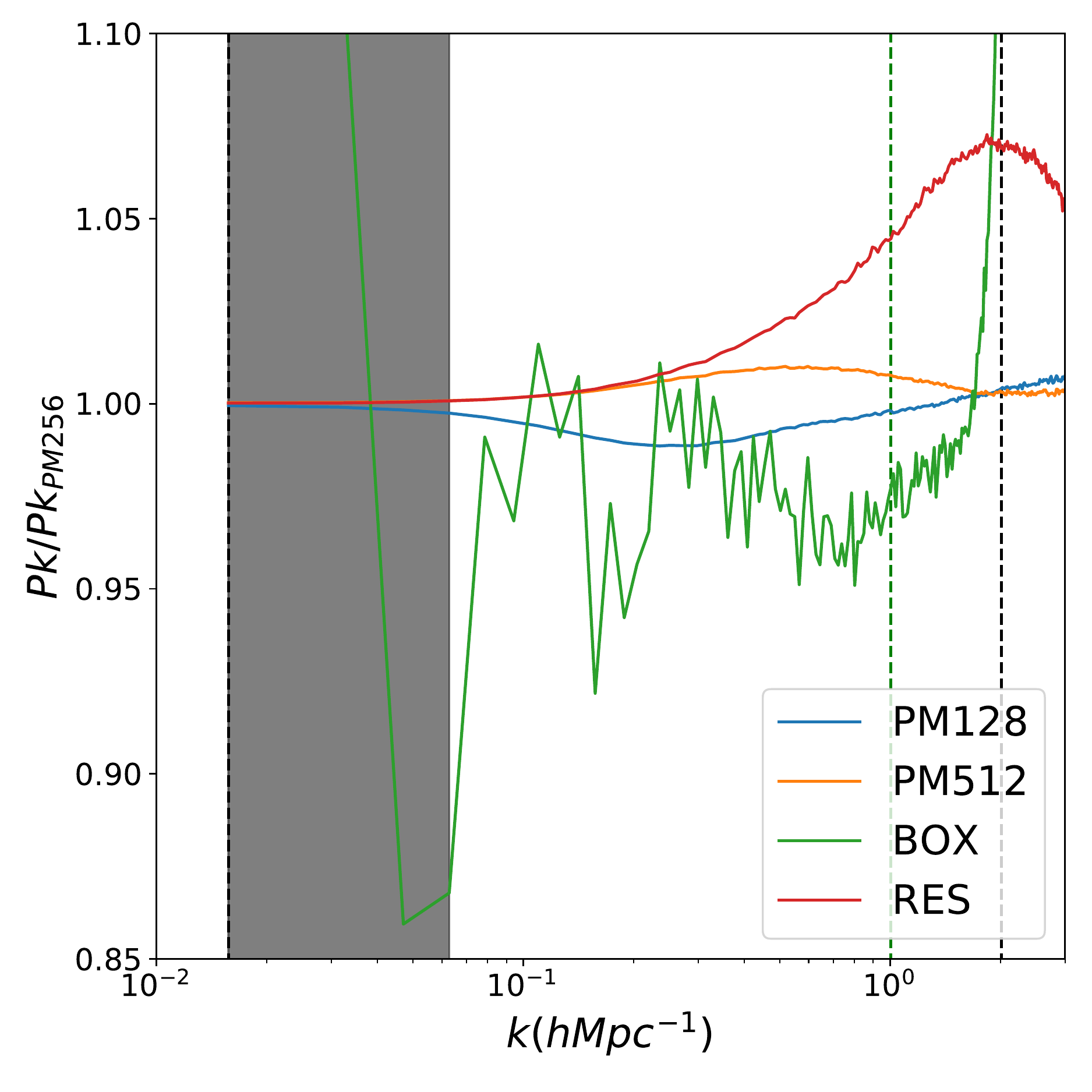}
\caption{Ratios of matter power spectra with respect to that of the PM256 simulation for the convergence runs at $z=0$. The blue (yellow, green, red) line shows the results from the PM128 (PM512, BOX, RES) run. The black dashed lines show the box size limit and resolution limit of our PM128, PM256, PM256 and RES simulations. The green dashed line shows the resolution limit of our BOX simulation. The difference between different number of grids is within $1\%$, which is accurate enough. The number of grids has negligible effects on our ME-Gadget code. The box size effect and resolution effect are $<5\%$ in our interested scale $k<1h\mathrm{Mpc}^{-1}$. The cosmic variance is the major reason of the difference between BOX and other runs, so we shaded the $k$ range close to the box size limit, where the difference is not due to simulation itself. The simulation setting of PM256 is enough for us to capture the physical insights from the non-linear evolution of the large scale structure.}\label{fig:convergence}
\end{figure}

In this section, we show that it is reasonable to modify the particle-mesh gravity only, and our code can be extended self-consistently to larger box sizes or higher resolutions. The simulations we performed for these tests use the parameters of Model I PBSH set listed in Table~\ref{tab.parameters}. This set of parameters has the largest interaction strength among all four models, and consequently leads to the largest deviation from $\Lambda$CDM. As shown in Fig.~\ref{fig:convergence}, changing the number of grids affects the non-linear matter power spectrum at $z=0$ by at most $1\%$. The modification of particle-mesh gravity on the grids represents the DE perturbation. We find that the influence of number of grids is minor as long as the number is enough for capturing the DE perturbation at large scales. We note that an accuracy of $\lesssim1\%$ has been found to be sufficient for the next generation surveys \cite{aemulus1}. 

\begin{figure}
\includegraphics[width=0.5\textwidth]{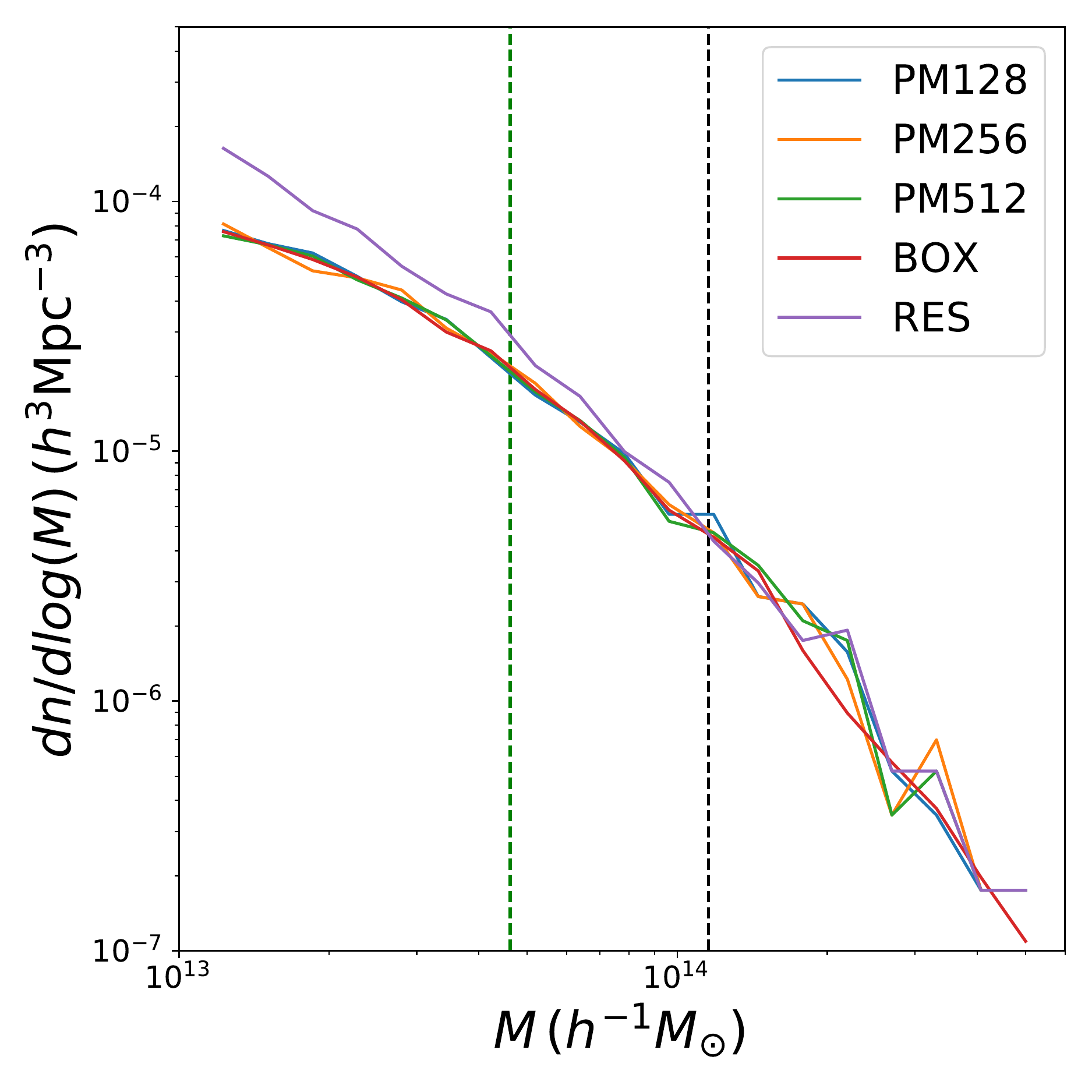}
\caption{Halo mass function of the convergence test results. Green (Black) dashed line shows the halo with 200 (500) particles. Resolution plays a major role in halos with particles less than 500, but not significant for halos with particles more than 500.}\label{fig:hmf_z0_con}
\end{figure}

The box size mainly affects the power spectrum at large scales. As we can see from Fig.~\ref{fig:convergence},  the difference is mainly caused by the cosmic variance. On the other hand, the resolution mainly affects the power spectrum close to the Nyquist limit at small scales. The effects of box size and resolution only introduce $\lesssim5\%$ difference in the range we are interested i.e., $k<1h\mathrm{Mpc}^{-1}$. Figure \ref{fig:hmf_z0_con} shows the halo mass functions at $z=0$ for different convergence test runs. It is clear that the resolution plays an important role in the halo mass function at the low mass end. For halos with more than 500 particles however, such effect is negligible. Since we use finite number of particles to represent the dark matter fluid and trace the evolution, the corresponding systematic bias is inevitable. We find that this bias is consistent with the $\Lambda$CDM simulations with Gadget2 as discussed in \cite{aemulus1}. Thus the systematic uncertainty in our code is at the same level as the original Gadget2 code. Therefore, we are confident about our simulation results. In the scientific runs, we use $256^3$ particles and $256$ grids per dimension within a $400h^{-1}$Mpc box. This choice of parameters passes the convergence test and is a balance between accuracy and computation costs. In addition, we conclude that our code is accurate and efficient to be further extended to future larger simulations.

\section{Results}\label{sec:result}

We have performed nine sets of scientific simulations that can be classified into three groups; a) one run for $\Lambda$CDM model, b) four runs for IDE models constrained by Planck alone (PC here after), c) four runs for IDE models constrained by Planck+Bao+SNIa+H0 (FC here after). From Table~\ref{tab.parameters}, we can see that in IDE\_I, the interaction parameter $\xi_2$ is constrained to be smaller than zero. It means in IDE\_I, energy is transferred from DM to DE. In IDE\_II, III and IV, the interaction parameters are all constrained to be larger than zero, which means that there is energy flow from DE to DM.  

Due to lack of N-body simulation for IDE models, an initial attempt to get the low redshift non-linear matter power spectrum was proposed in \citep{An2018}. This was done by adding the non-linear correction, so-called halofit \citep{takahashi2012halofit}, onto the linear matter power spectrum in IDE models. This is an approximate treatment, since it is only true when the IDE model does not deviate much from the $\Lambda$CDM model. Because the halofit is an empirical fit to $\Lambda$CDM model N-body simulations, it cannot be directly applied to IDE models, especially when the interaction parameter is large enough.  Fully self-consistent simulation pipeline is called for to explore the physics on non-linear structure formation when there is interaction between dark sectors at low redshifts.

\begin{figure*}
\includegraphics[width=\textwidth]{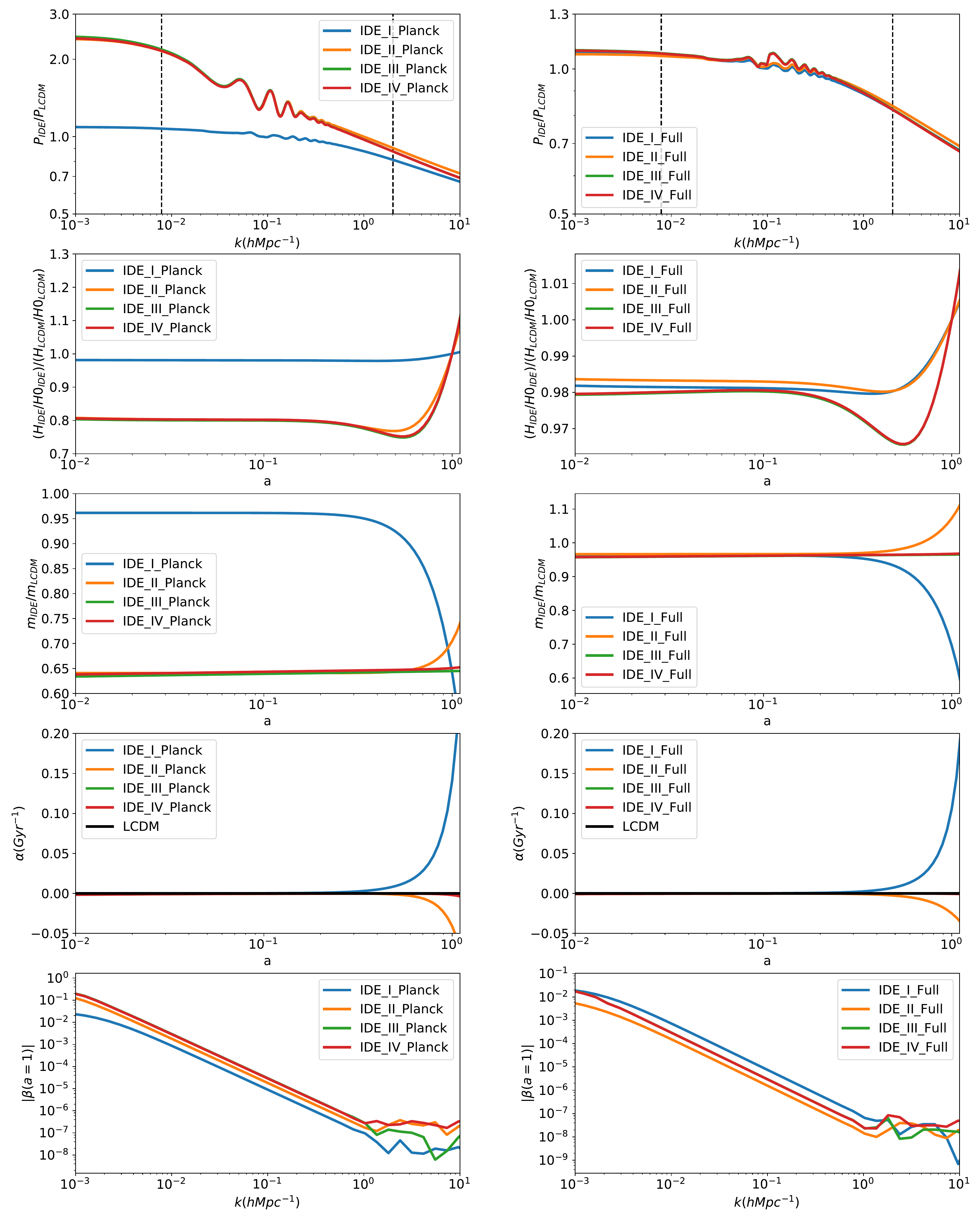}

\caption{Interpolation tables we used in the simulations. PC simulations are shown on the left while FC simulations are given on the right. From top to bottom, they are initial matter power spectrum ratio between IDE models and the $\Lambda$CDM model (denote as LCDM in the plot) at $z=49$, H/H0 ratio, simulation particle mass ratio, drag force $\alpha(a)$ and fifth force $\beta(a=1,k)$. Together with ME-Gadget code and these tables, one can reproduce all the scientific simulations we have shown in this paper.}\label{fig:interpolation}
%
%
\end{figure*}

We plot the interpolation tables used in the simulations in Fig.~\ref{fig:interpolation}. We notice that IDE\_I and IDE\_II models are different from $\Lambda$CDM both in FC parameter sets and PC parameter sets. Comparing with the $\Lambda$CDM model, it is clear that the particle masses in IDE\_I (IDE\_II) get lower (higher), while the velocity dependent accelerations get larger (smaller) quickly at low redshifts. These differences are caused by the energy flow from DM to DE in IDE\_I (DE to DM in IDE\_II) which increases sharply at low redshifts. Thus, we expect to see significant differences in large scale structures at low redshifts in IDE\_I and IDE\_II simulations.

\subsection{Density Field}
\begin{figure}
\includegraphics[width=0.5\textwidth]{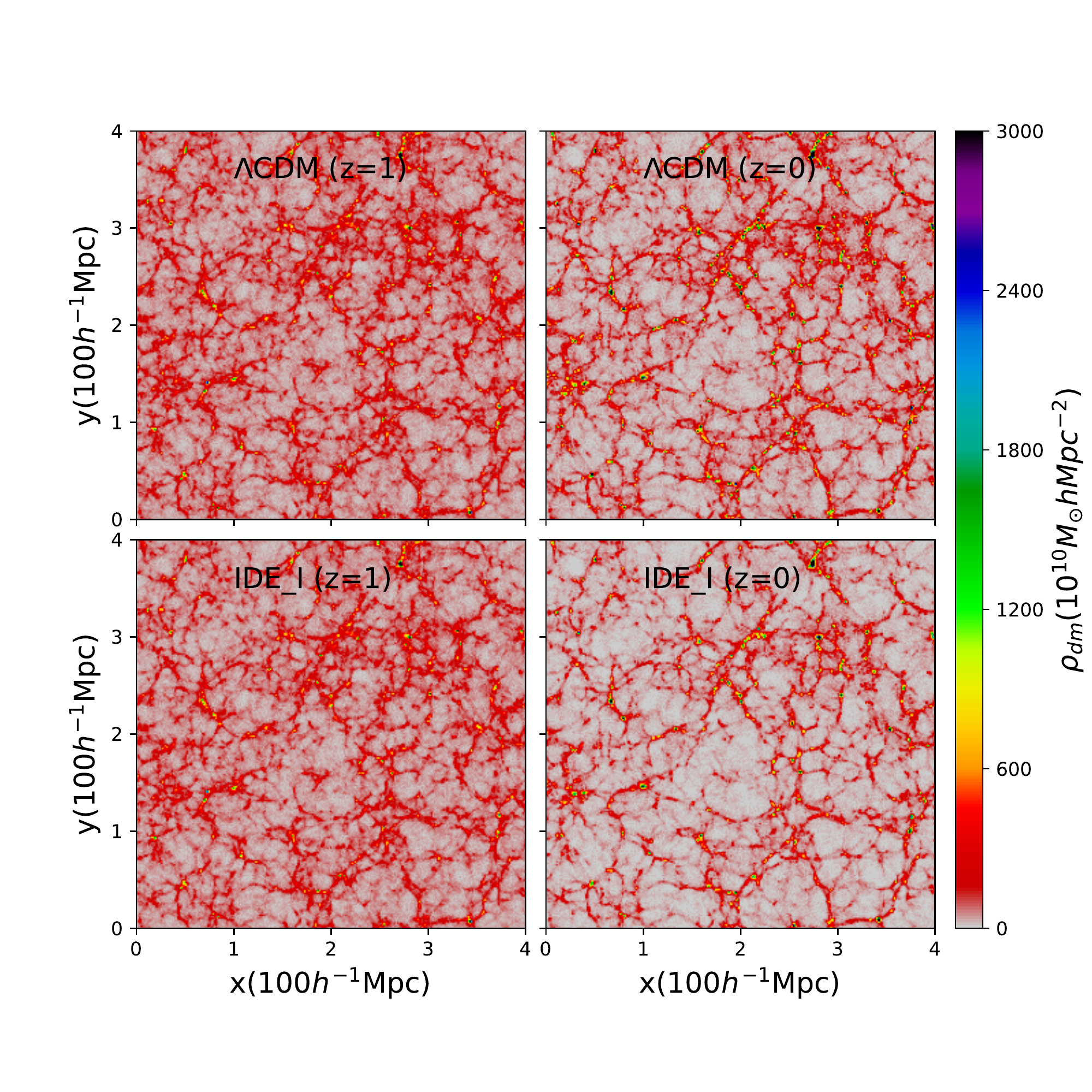}
\caption{Density distribution comparison between FC IDE\_I simulation and $\Lambda$CDM simulation. Upper left (right) panel shows the density distribution of $\Lambda$CDM simulation at $z=1$ ($z=0$.) Lower left (right) panel shows the density distribution of FC IDE\_I simulation at $z=1$ ($z=0$). The density distribution is plotted over a slice of the simulation box with $10h^{-1}$Mpc thick. The color denotes the surface density in $10^{10}hM_{\odot}\mathrm{Mpc}^{-2}$. There is no significant difference at $z=1$, but obviously, IDE\_I becomes less dense at $z=0$ due to the transfer from dark matter to dark energy.}\label{fig:slice}
%
%
\end{figure}
Figure \ref{fig:slice} shows the projected matter density distribution in a slice with a thickness of $10h^{-1}$Mpc. Comparing the FC IDE\_I simulation and the $\Lambda$CDM model, it is clear that IDE\_I structure is less dense than that in the $\Lambda$CDM. This is caused by the quick flow of energy from DM to DE at low redshifts. The speed of the energy flow in IDE\_I is proportional to the DE average density, which decreases very slowly ($\propto\sim a^{-0.3}$) compared to that of DM ($\propto a^{-3}$). Thus, DM flowing into DE accelerates in terms of scale factor, when the universe becomes DE dominated at low redshifts. 
\begin{figure}
\includegraphics[width=0.5\textwidth]{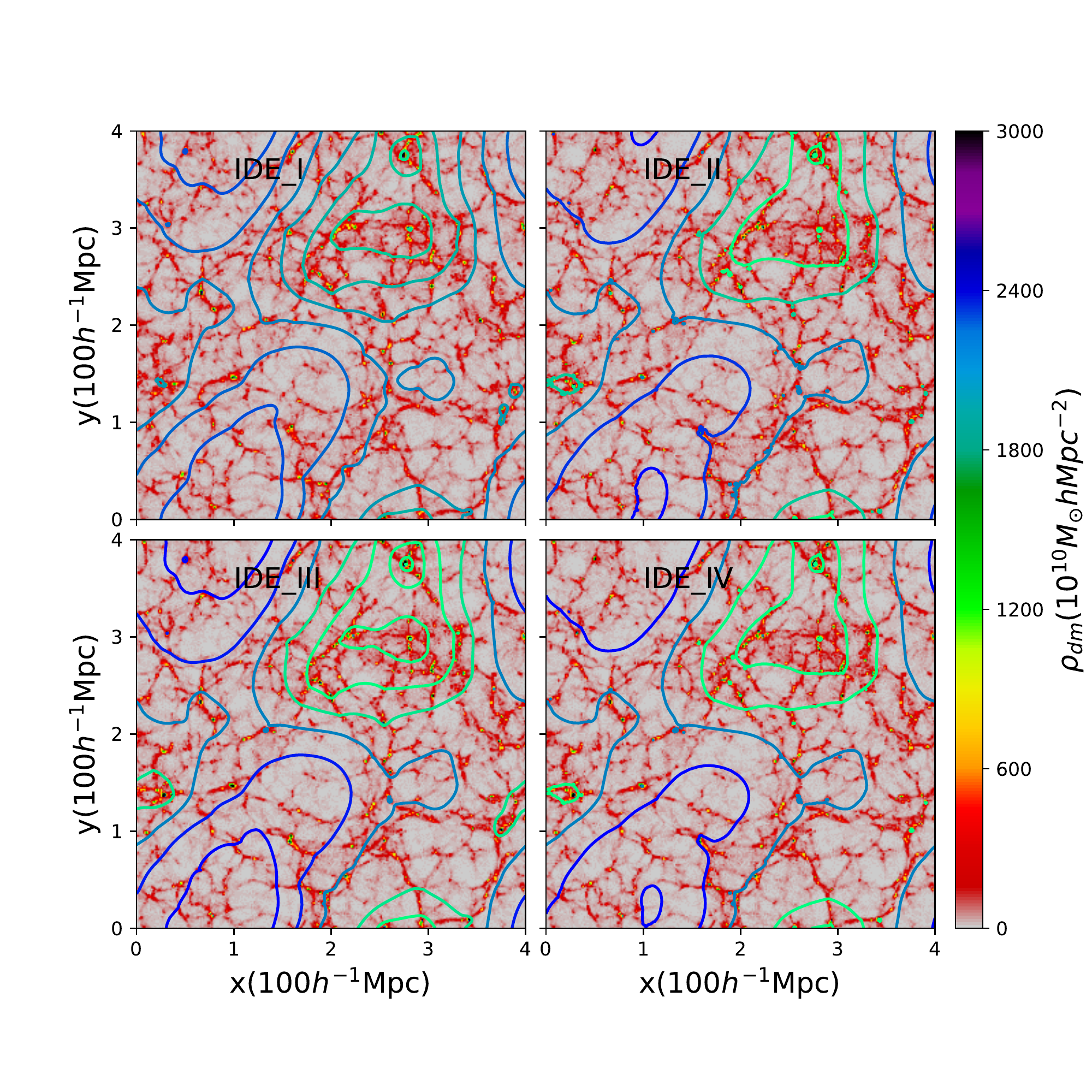}
\caption{The dark matter density distributions are shown in reddish colors and the dark energy distributions are shown in the contours for four PC simulations. The colors of the contour represent the DE density, blue (green) contours represent low (high) DE density. The DM density distribution share the same scale and unit with Fig.~\ref{fig:slice}, the dark energy contour is in arbitrary unit, but the same for these four plots, just to illustrate that the dark energy perturbation is only effective on large scales.} \label{fig:deslice}
\end{figure}
\begin{figure}
\includegraphics[width=0.5\textwidth]{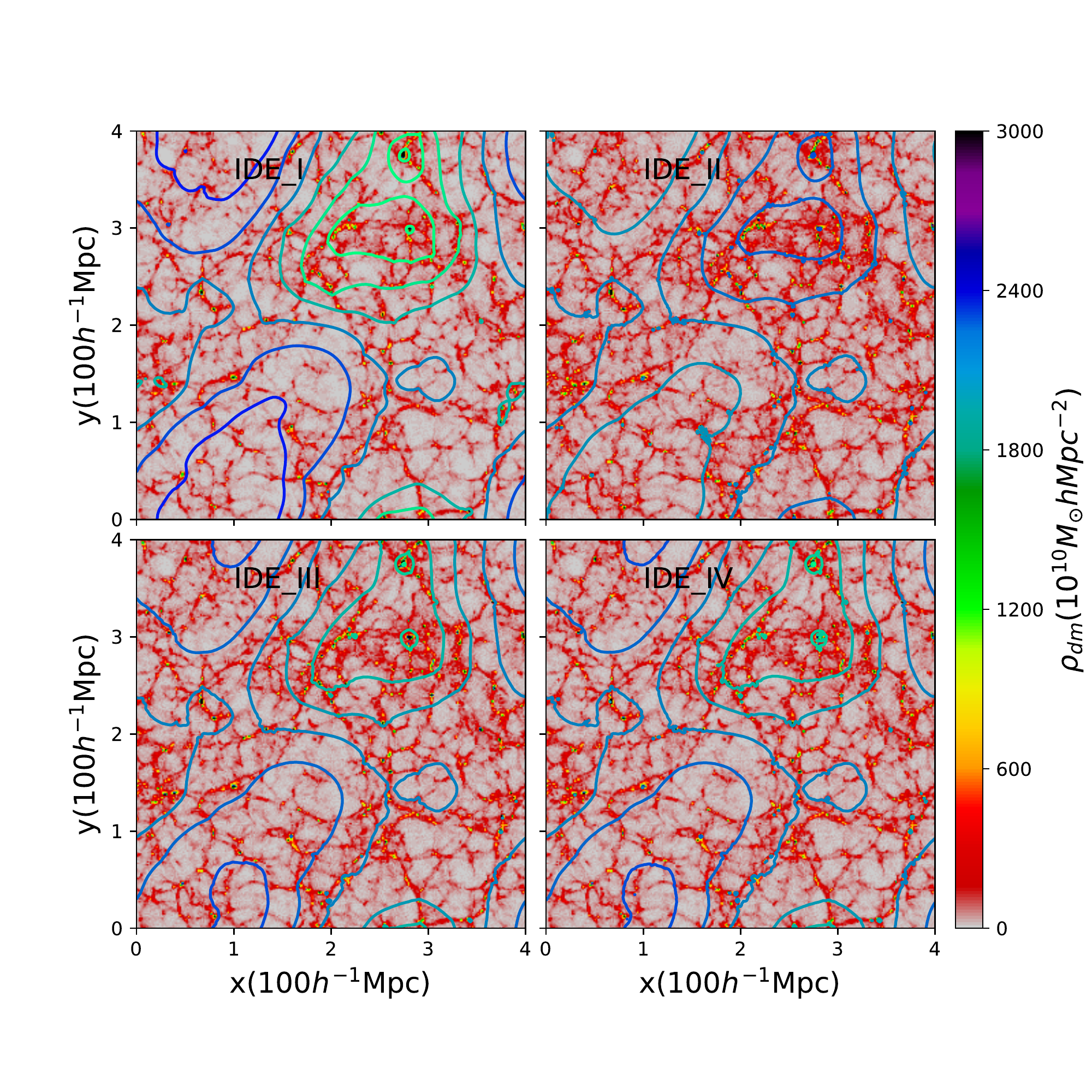}
\caption{The dark matter density distributions are shown in reddish colors and the dark energy distributions are shown in the contours for four FC simulations. The plotting sets are similar to Fig.~\ref{fig:deslice}.} \label{fig:deslice2}
\end{figure}

\begin{figure}

\includegraphics[width=0.5\textwidth]{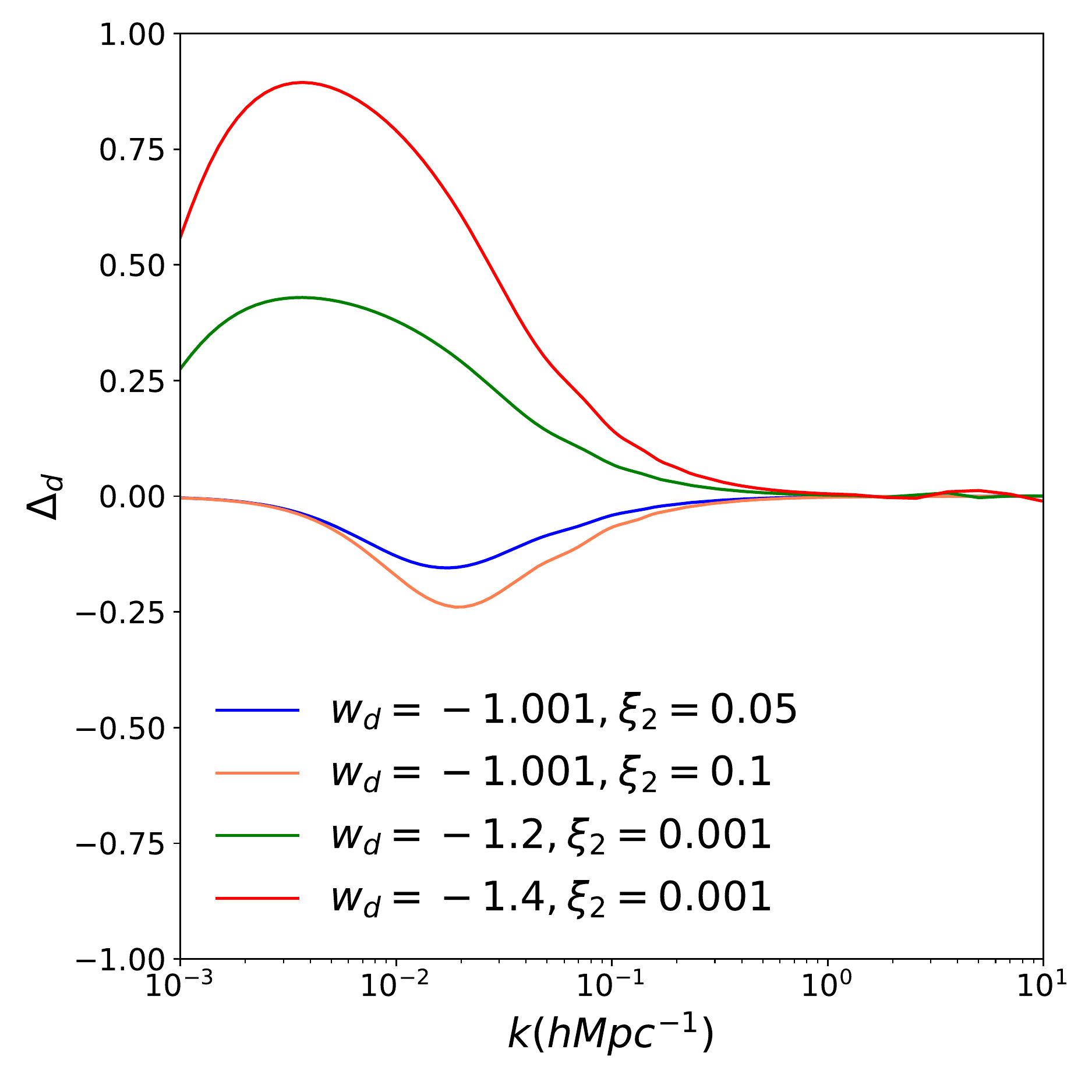}
\caption{The dark energy perturbations $\Delta_d$ at $z=0$ for different $\omega_d$ and $\xi_2$ are shown here. $\Delta_d>0$ means dark energy follows the dark matter distribution, while $\Delta_d<0$ means dark energy follows the dark matter distribution inversely. More deviation from $\omega_d=-1$ leads to more clustering and larger $\xi_2$ leads to more anti-clustering.}\label{fig:delta_de}
\end{figure}

We show  DM and DE density distributions for four PC simulations in Fig.~\ref{fig:deslice} and FC simulations in Fig.~\ref{fig:deslice2}. The DM density distributions at large scale are similar for all four IDE models. Although the DE distribution is not homogeneous, its perturbation is much smaller than that of the DM, which is consistent with the linear investigations \cite{avelino2013prd}. The contours in Fig.~\ref{fig:deslice} and Fig.~\ref{fig:deslice2} showing the DE distribution were calculated by multiplying the DM density with $\beta(a,k)$ using linear calculation. In Fig.~\ref{fig:delta_de}, we show the dependence of DE perturbation on the DE equation of state $w_d$ and the interaction strength $\xi_2$. If $w_d=-1$, DE is the cosmological constant, there is no DE perturbation. If $w_d$ deviates from $-1$, DE perturbation grows as $\vert{w_d+1}\vert$, leading to stronger clustering together with DM. In contrast, as the interaction strength increases, DE perturbations become more negative leading to stronger anti-clustering with DM. 

In Fig.~\ref{fig:deslice} and Fig.~\ref{fig:deslice2}, the color of the contour represents the DE density. It is clear that, except for FC IDE\_II simulation, DE generally follows DM clustering, where DE is mostly concentrated in the most dense DM regions. There are however, DE condensations in the DM void regions. This can be understood by considering a wet sponge (DM) filled with liquid (DE) as an analogy. If the sponge is squeezed (DM collapse), part of the liquid (DE) will be compressed, while at the same time, there is also a part of the liquid that is squeezed out. This shows that although DE can participate in the structure formation, but it does not collapse exactly along with DM. This is consistent with the study of Layzer-Irvine equations in the linear formalism for the collapse of structure in the expanding universe \cite{He2010}. In sharp contrast, for FC IDE\_II simulation, DE is underdense in regions where DM of overdensity. This again can be understood by considering a wet sponge (DM) with incompressible liquid (DE), where most of the liquid (DE) is squeezed out instead of being compressed as the sponge squeezed (DM collapse).

\subsection{Matter Power Spectrum}
\begin{figure*}
\includegraphics[width=0.45\textwidth]{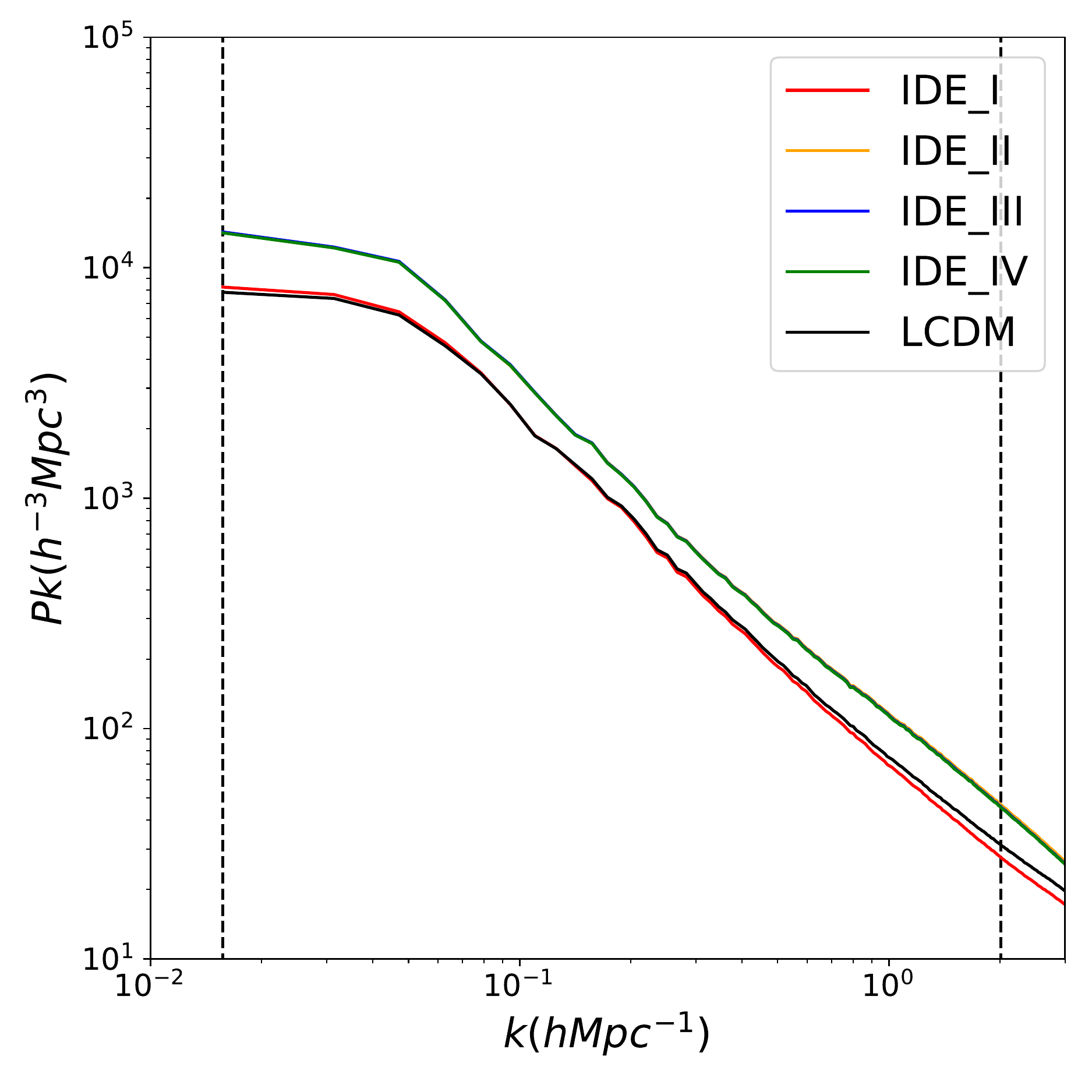}
\includegraphics[width=0.45\textwidth]{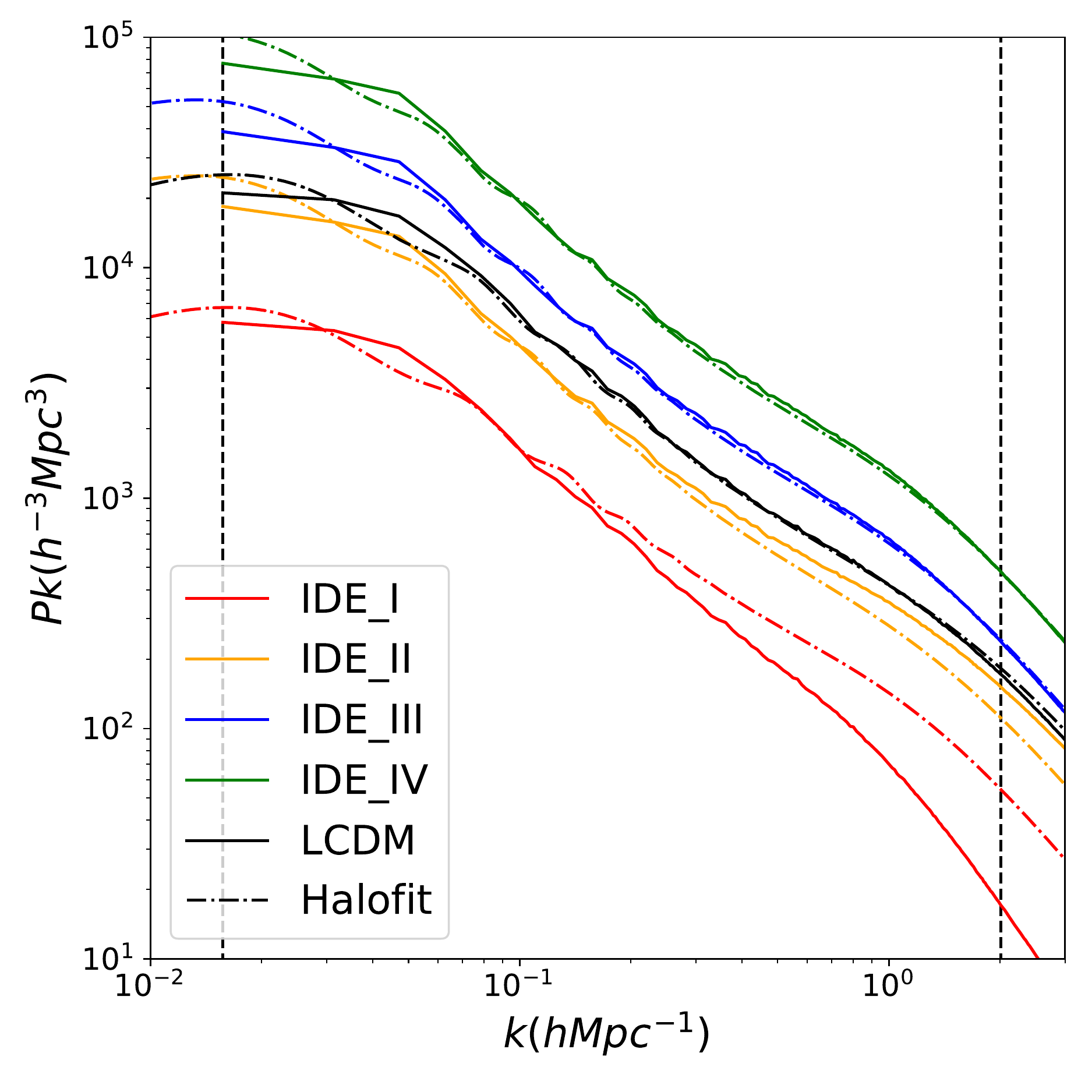}
\caption{The matter power spectra of PC simulations at $z=1$ ($z=0$) are shown on the left (right) panel. All models are similar to $\Lambda$CDM at $z=1$, except for some normalization difference. Notice that we have rescaled IDE\_I (IDE\_II, IDE\_III, IDE\_IV) by a factor of $\dfrac{1}{4}$ ($\dfrac{1}{2}, 1, 2$) for better illustration at $z=0$. The slope of matter power spectrum of IDE\_I and IDE\_II are clearly different with the other models at $z=0$. The calculated non-linear matter power spectra by \cite{takahashi2012halofit} are given in dash-dotted lines. $\Lambda$CDM model, IDE\_III and IDE\_IV can be well represented by \cite{takahashi2012halofit}, but it fails for IDE\_I and IDE\_II due to the non-trivial non-linear evolution of these two models. }\label{fig:pk}
\end{figure*}

\begin{figure*}
\includegraphics[width=0.45\textwidth]{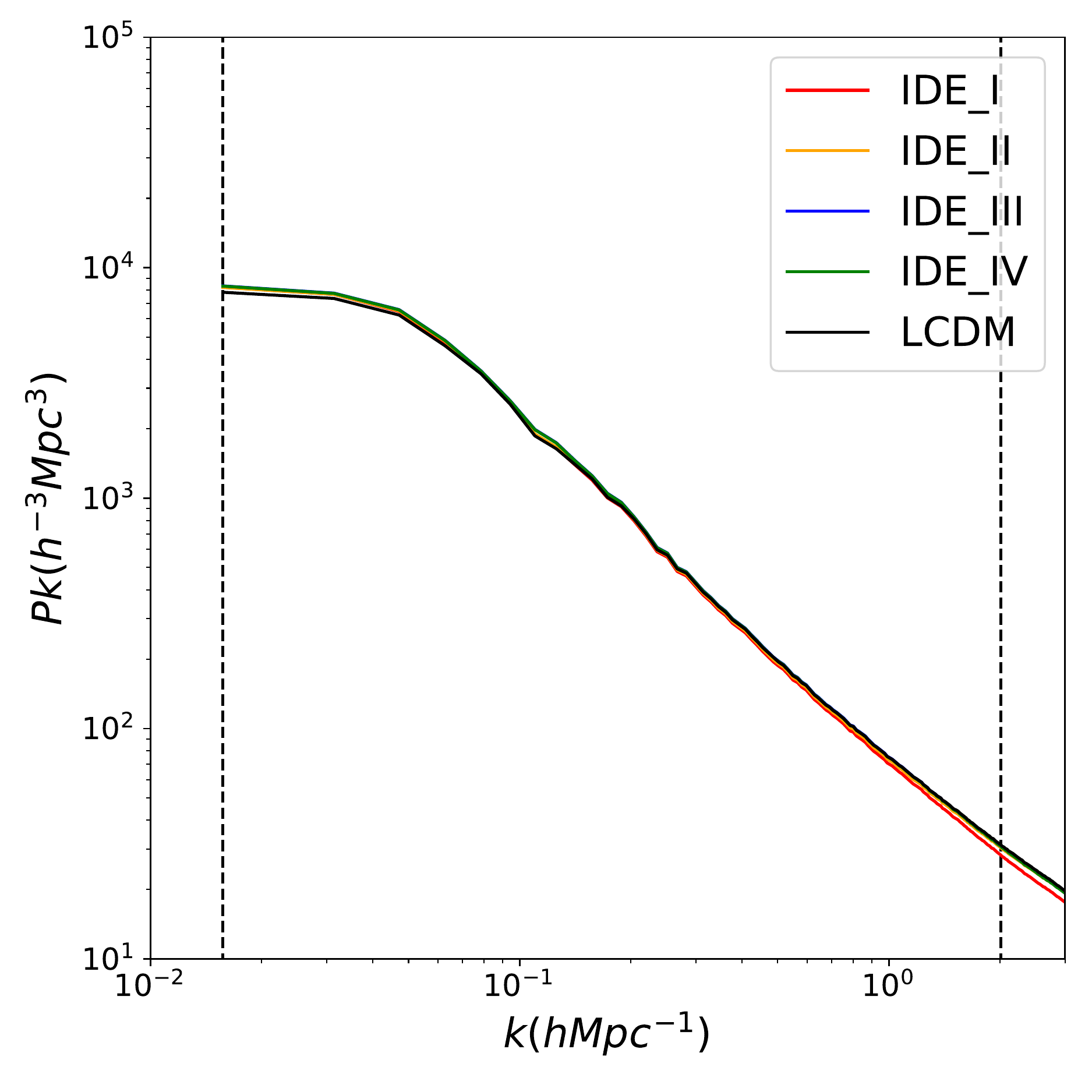}
\includegraphics[width=0.45\textwidth]{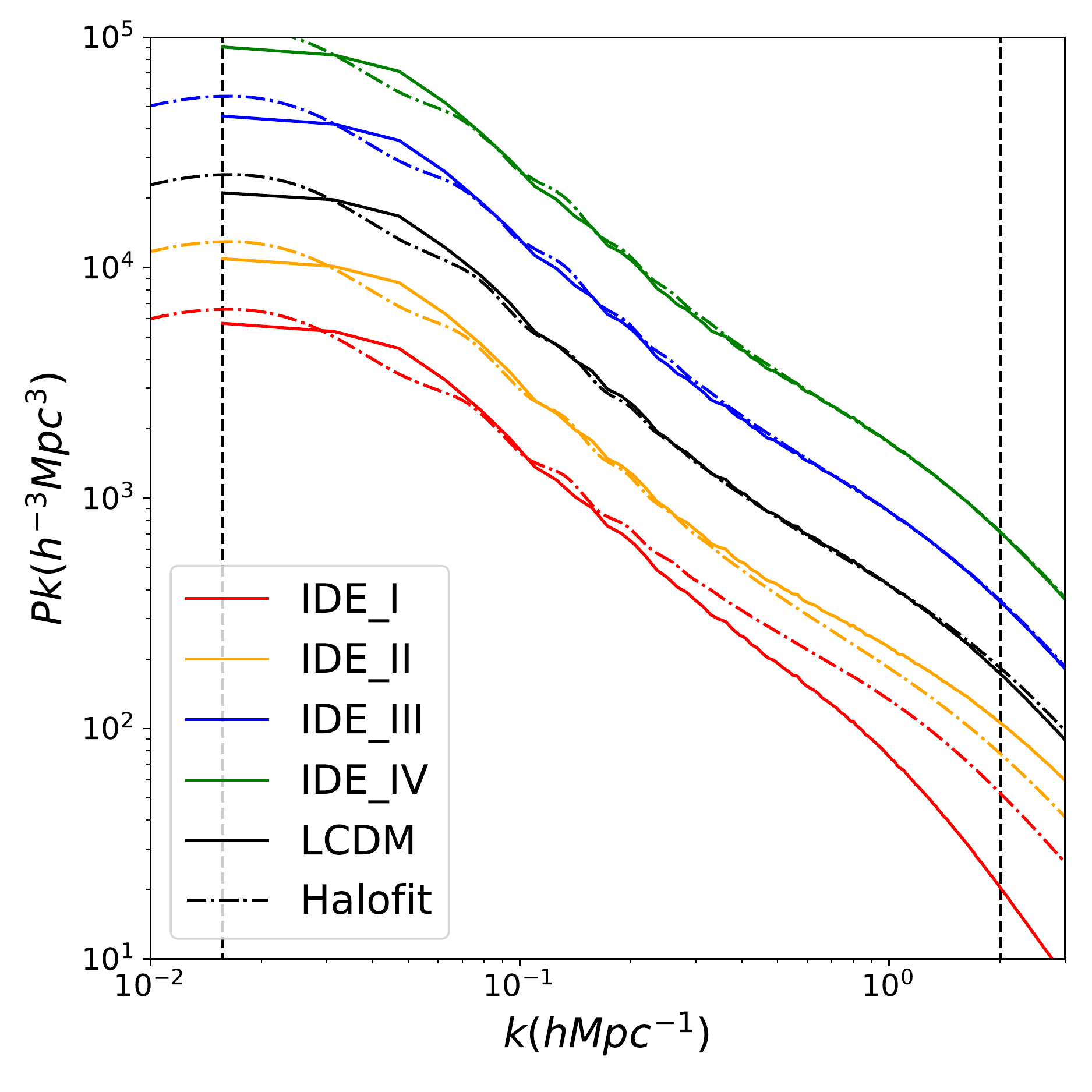}
\caption{Similar plot to Fig.~\ref{fig:pk}, but show the matter power spectra of FC models. All matter power spectra of different models are almost identical to $\Lambda$CDM model at $z=1$, except for IDE\_I with some minor difference. Notice that we have rescaled IDE\_I (IDE\_II, IDE\_III, IDE\_IV) by a factor of $\dfrac{1}{4}$ ($\dfrac{1}{2}, 2, 4$) for better illustration at $z=0$. At $z=0$, IDE\_I and IDE\_II are clearly different with the other three. IDE\_III and IDE\_IV keep identical to $\Lambda$CDM at $z=0$. The FC models are better normalized to $\Lambda$CDM model than the PC models, without rescaling, all five curves are almost identical at large scales.}\label{fig:pk2}
\end{figure*}

We measure the matter power spectra of the simulations using ComputePk code\citep{computepk}. We show the matter power spectra for the PC simulations in Fig.~\ref{fig:pk} and the FC simulations in Fig.~\ref{fig:pk2}. We find that at $z=1$, the matter power spectra of all the IDE models are similar to that of the $\Lambda$CDM model, except for some normalization differences in the PC simulations leading to some overall offsets. However, at $z=0$, the matter power spectra of IDE\_I and IDE\_II are clearly different from those of IDE\_III, IDE\_IV and $\Lambda$CDM. It is clear that the matter power spectrum of IDE\_I is suppressed with a steeper slope at $k>0.1h$Mpc$^{-1}$ than other models. In contrast, the matter power spectrum of IDE\_II is enhanced at $k>0.8h$Mpc$^{-1}$ with a shallower slope compared to $\Lambda$CDM model. This can easily be attributed to the direction of the energy flow. In the constrained IDE\_I, energy flows from DM to DE, while for IDE\_II the energy transfers in the opposite direction. For IDE\_III and IDE\_IV models, we find that the matter power spectra from the FC simulations are very similar to that of the $\Lambda$CDM model.   

We also compare the matter power spectra measured from our simulations with the halofit non-linear power spectrum \citep{takahashi2012halofit}. The purpose of doing so is to check the validity of employing the halofit model to calculate nonlinear corrections adopted in \cite{An2018}. For models whose matter power spectra are not much different from that of the $\Lambda$CDM model, such as IDE\_III and IDE\_IV, it is safe to use halofit. However, for models with clearly different matter power spectra from that of the $\Lambda$CDM model i.e., IDE\_I and IDE\_II, using halofit blindly can lead to meaningless and wrong results. Because the non-linear evolution in these models is highly non-trivial, and it is drastically different from  the $\Lambda$CDM model. In short, halofit should not be used as a simplification without self-consistent analysis and simulation. 

\subsection{Halo Mass Function}
\begin{figure*}
\includegraphics[width=0.45\textwidth]{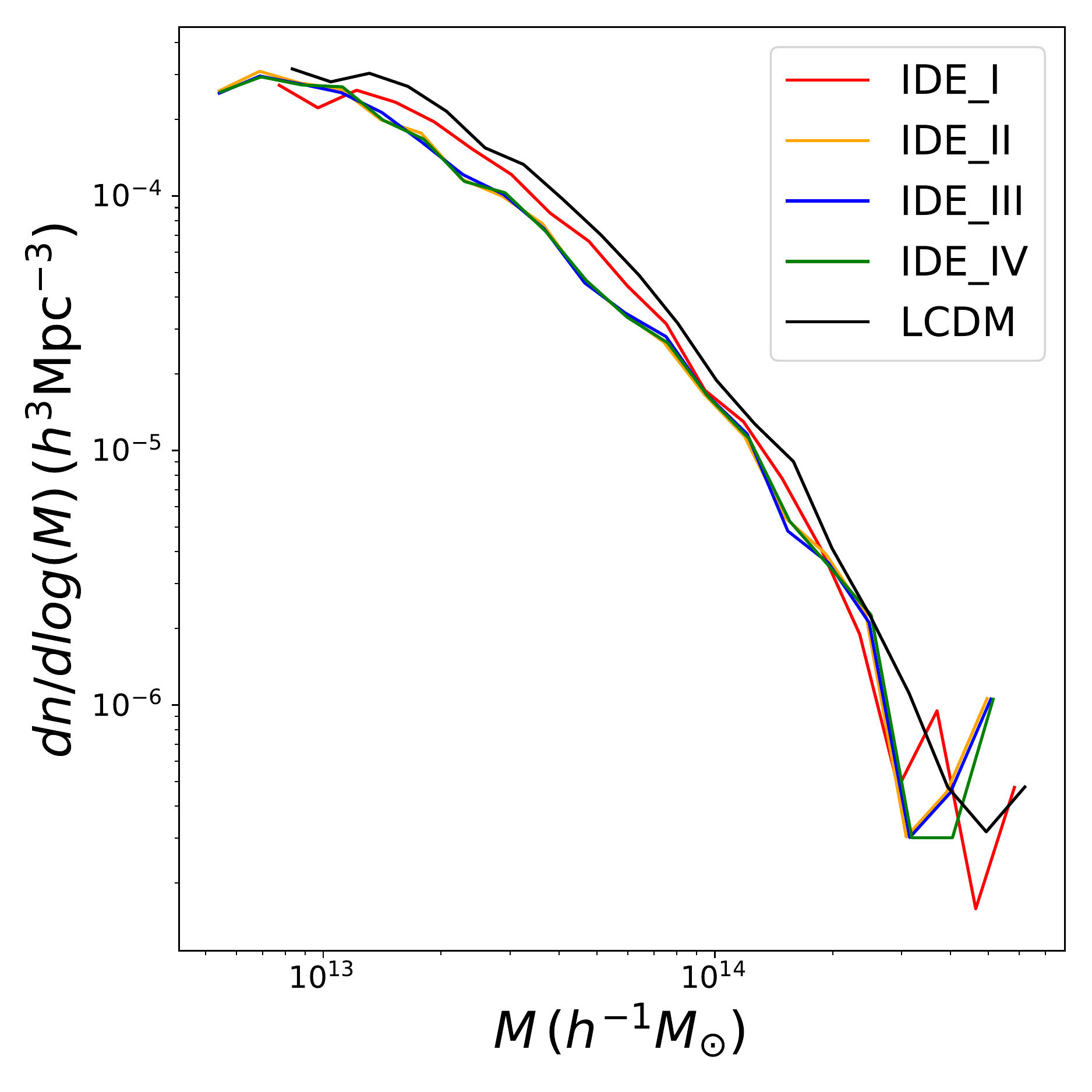}
\includegraphics[width=0.45\textwidth]{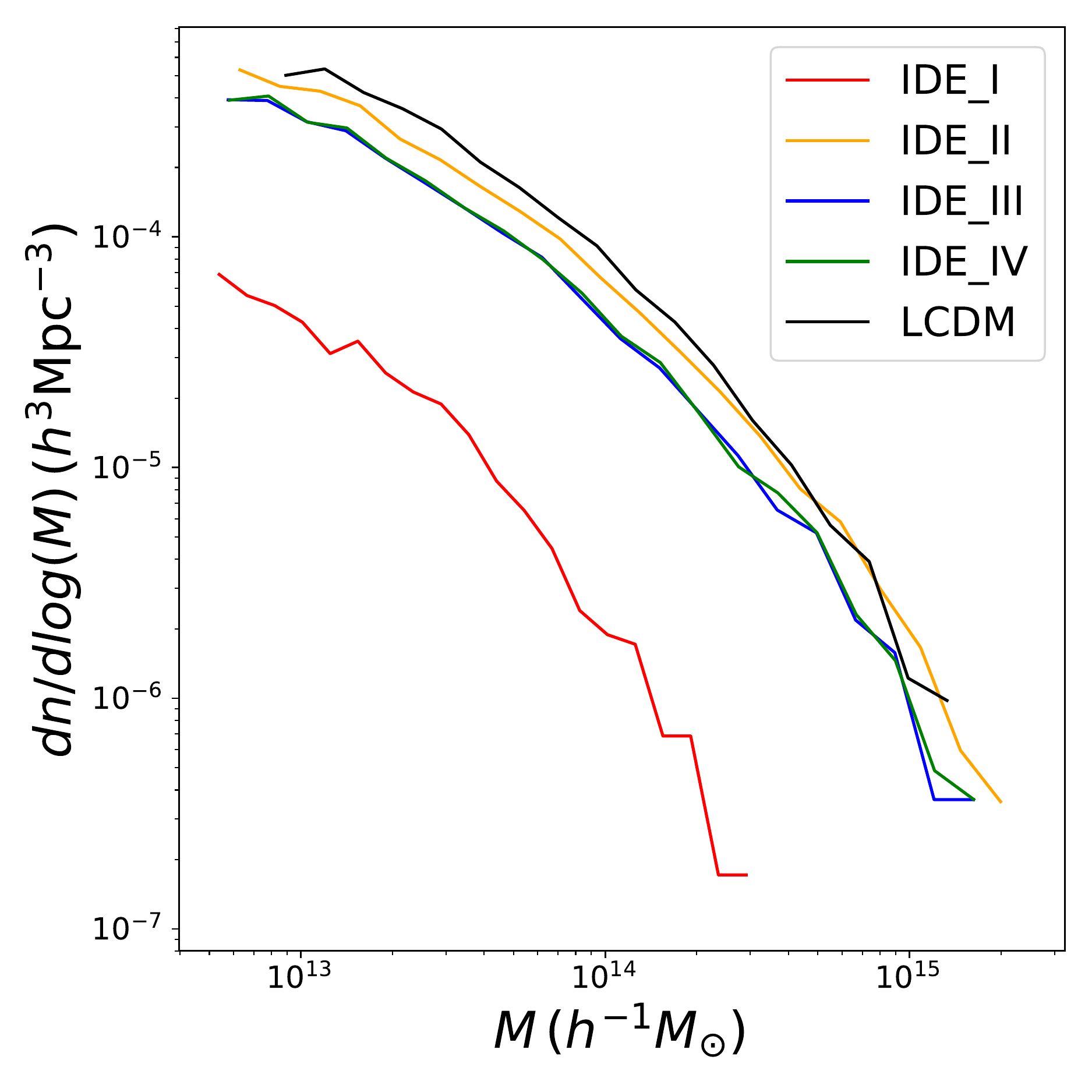}
\caption{The halo mass functions of PC simulations at $z=1$ ($z=0$) are shown on the left (right) panel. Notice that due to the energy flow from DM to DE, the red line, representing IDE\_I, is much lower than the others at $z=0$. The amplitudes of the models are different due to the different normalizations given by the PC parameters.}\label{fig:hmf}
\end{figure*}

\begin{figure*}
\includegraphics[width=0.45\textwidth]{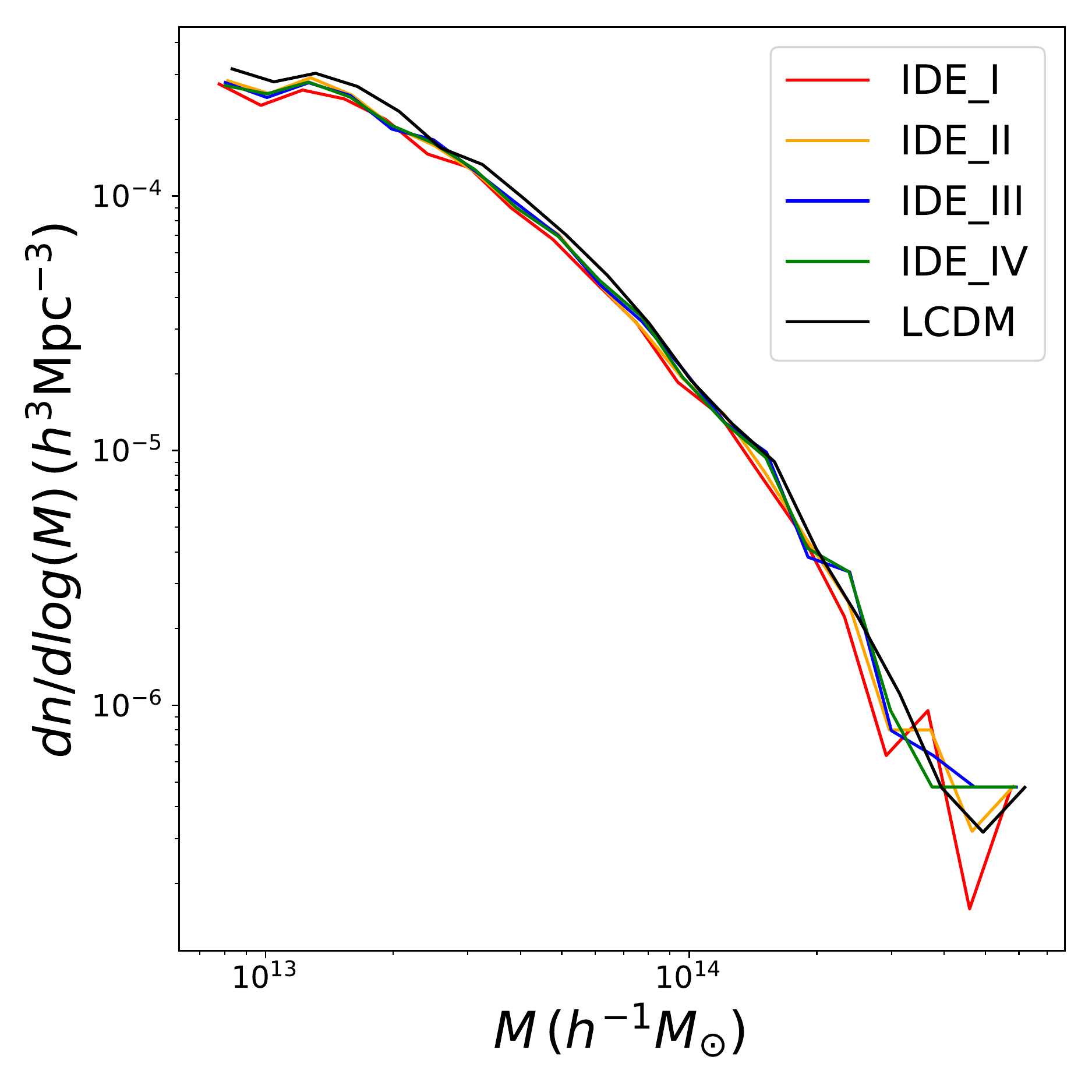}
\includegraphics[width=0.45\textwidth]{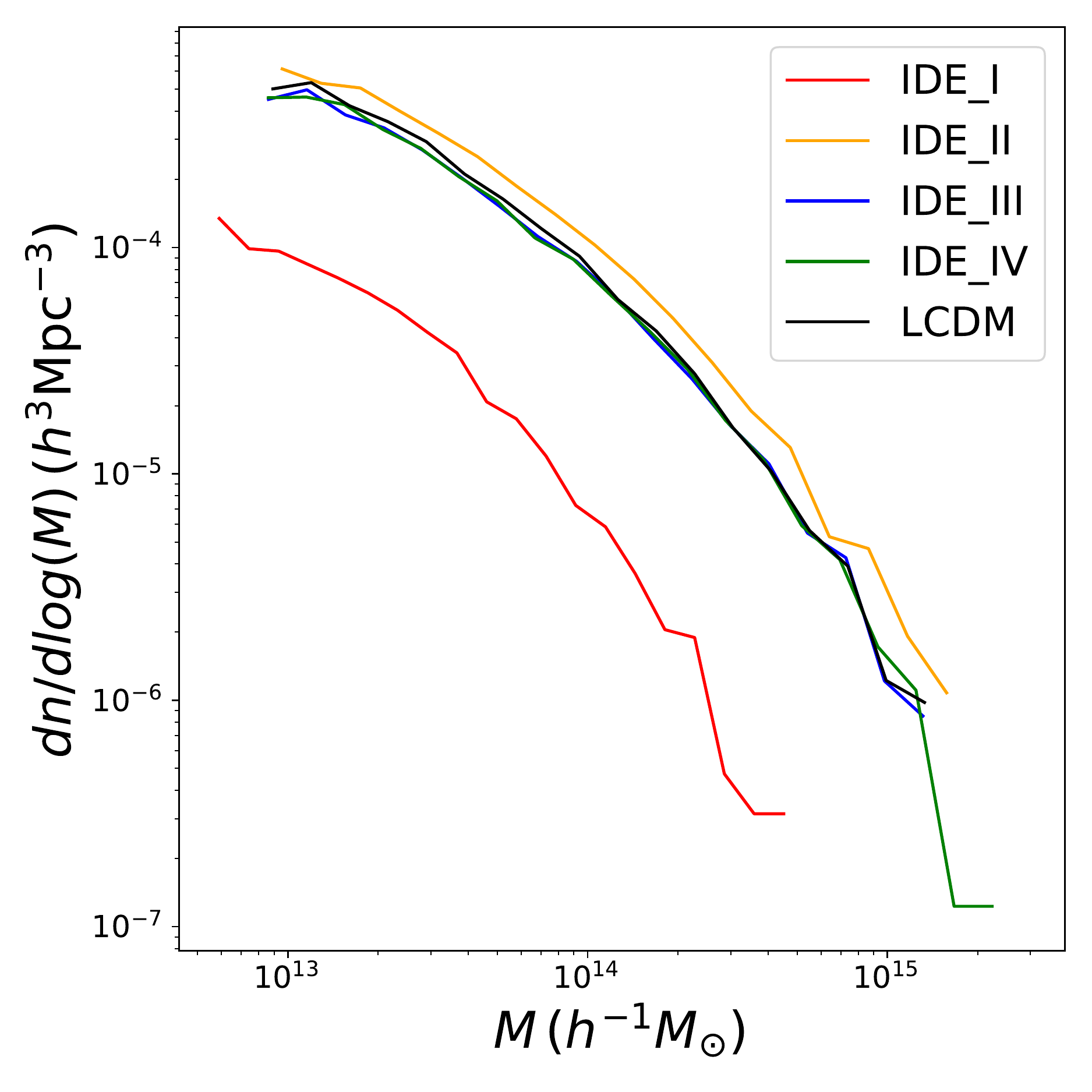}
\caption{Similar plot to Fig.~\ref{fig:hmf}, but show the halo mass functions of FC simulations. After proper normalization with the FC parameters, IDE\_III, IV and $\Lambda$CDM, at both $z=1$ and $z=0$, are not distinguishable. We clearly see the number of halos in IDE\_I is much less than the other models at $z=0$.}\label{fig:hmf2}
\end{figure*}

We identify halos with an overdensity parameter $\Delta_{200}=200$ with respect to the mean background density  using AHF\citep{ahf}, and measure the halo mass functions in our simulations. The halo mass functions of PC simulations are shown in Fig.~\ref{fig:hmf}, while those of FC simulations are shown in Fig.~\ref{fig:hmf2}.
It can be seen from Fig.~\ref{fig:hmf} and Fig.~\ref{fig:hmf2} that the halo mass function in IDE\_I from $z=1$ to $z=0$ compared to $\Lambda$CDM model changes dramatically. For IDE\_II, there is also a noticeable difference. Such differences compared to the $\Lambda$CDM model can be used to provide strong constraints by our further analysis in the future. It is worth noting that using N-body simulations, we are in a position to constrain IDE models better than previous linear level studies using PBSH \citep{costa2017jcap}. This shows the benefit of performing self-consistent simulations in studying non-linear structure formation in IDE models.  

\section{Summary and Discussion}\label{sec:conclusion}
We have devised a fully self-consistent simulation pipeline for IDE models, the core of which is the novel N-body simulation code. We have modified Gadget2 \citep{gadget2} so that it can accept arbitrary inputs including Hubble diagram, simulation particle mass, velocity dependent acceleration and DE perturbation. This modified code is called ME-Gadget. We use ME-Gadget to simulate the non-linear evolution of IDE models. This idea was first suggested by \cite{baldi2010mnras}. However, they only considered a specific DE model with constant perturbation and adopted the initial conditions from the $\Lambda$CDM model. Both of these are major drawbacks which we have successfully overcome in our pipeline. We performed comparison and convergence tests and found that our pipeline is accurate as well as efficient. We have tested the effect of neglecting DE perturbation at small scales by varying the number of mesh grids used to calculate the gravity. We show that the effect is less than $1\%$ if the number of mesh grids is enough to cover the main effective DE perturbation scales. We have also tested that the effects of different box sizes and resolutions are less than $5\%$ in the matter power spectrum. Thus, we have successfully developed a fully self-consistent pipeline for simulating IDE models which includes a) simulating the effect of DE perturbation at large scales,  b) generating the pre-initial conditions, c) using second-order Lagrangian Perturbation Theory consistently, and d) employing the CAMB code to generate the initial matter power spectrum.

Using the cosmological parameters constrained by \cite{costa2017jcap}, we performed nine sets of scientific simulations applying our pipeline. These parameters passed the constraints from Planck CMB observation and Planck+BAO+SNIa+H0 combined measurements. IDE\_I and IDE\_II, whose interactions between DM and DE are proportional to the energy density of DE, show significant difference between the direct simulations and the prediction from the naive halofit attempt. Since the non-linear matter power spectrum close to $z=0$ is powerful for studying non-standard cosmological models by comparing with observations, a self-consistent pipeline is indispensable. The significant differences between IDE\_I, IDE\_II and $\Lambda$CDM model indicate that tighter constraints can be put on these models by comparing the simulation results with observations at low redshifts. We have showed that simulations are vital for providing further constraints in the future using large-scale structures at low redshifts.

We summarize our results from the simulations in the following four points:
\begin{itemize}
\item[1]In general, if energy flows from DM into DE, the structure formation will be suppressed, and \textit{vice-versa}. However, the effect of the interaction between DM and DE on the non-linear evolution is non-trivial. Simulations are necessary for studying large scale structures for IDE models.
\item[2]If the interaction parameter is small, such as $\xi_1(\xi)\sim0.001$ in IDE\_III and IDE\_IV models, halofit can still be a good approximation. But if the interaction parameter is large, such as $\xi_2>0.03$ in IDE\_I and IDE\_II, halofit is not appropriate. 
\item[3]DE perturbations grow together with DM density perturbations, but at much larger scales of $\sim100h^{-1}$Mpc. The growth of DE perturbations depend on the equation of state and the interaction parameters. 
\item[4]Although allowed by combined Planck, BAO, SNIa and H0 observations, the results of simulations of IDE\_I and IDE\_II models are significantly different from the $\Lambda$CDM model in nonlinear structure formation at $z=0$. This indicates that low redshift observations can be a powerful tool for refining IDE models in the future. However for IDE\_III and IDE\_IV models, we cannot count on the nonlinear simulation to  distinguish them from the $\Lambda$CDM model. 
\end{itemize}

In the future, we plan to use the ME-Gadget code to perform multiple simulations with larger box sizes and higher mass resolutions, and cover larger parameter space to build up emulators for the observable. We will use our simulations to put further constraints on IDE models using observations of large scale structures at low redshifts. We forecast a large improvement in the constraints for the IDE\_I and IDE\_II models. Further studies will also be done for IDE models with quantum field theory origin. 

\section*{Acknowledgement}
We thank Projjwal Banerjee for useful comments.
J. Z acknowledges the support from China Postdoctoral Science Foundation 2018M632097.
W. L acknowledges the support from NSFC 11503064, Shanghai Jiao Tong University
and University of Michigan Joint Fundation (AF0720054) and the support from WPI Japan. The work of B. W was partially supported by NNSFC.

\bibliographystyle{apsrev4-1}
\bibliography{DMDEI}

\end{document}